\documentclass[twocolumn,final]{IEEEtran}
\usepackage{amsmath,graphicx}
\usepackage{amssymb}
\usepackage{amscd}
\usepackage{multirow}
\usepackage{subcaption}
\usepackage{pbox}
\usepackage{cite}
\usepackage{algorithmic}
\usepackage{algorithm}

\def\expk{{e^{-i \frac{2\pi}{T}k t}}}

\title{Sub-Nyquist Sampling and Fourier Domain Beamforming in Volumetric Ultrasound Imaging}
\author{\IEEEauthorblockN{Amir Burshtein\IEEEauthorrefmark{1},
Michael Birk\IEEEauthorrefmark{1},
Tanya Chernyakova\IEEEauthorrefmark{1},
Alon Eilam\IEEEauthorrefmark{1},
Arcady Kempinski\IEEEauthorrefmark{7} and
Yonina C. Eldar\IEEEauthorrefmark{1}}
\IEEEauthorblockA{\IEEEauthorrefmark{1}Department of Electrical Engineering, Technion, Israel Institute of Technology, Haifa, Israel}

\IEEEauthorblockA{\IEEEauthorrefmark{7}GE Healthcare, Haifa, Israel}}

\begin{document}
%
\maketitle
\begin{abstract}
One of the key steps in ultrasound image formation is digital beamforming of signals sampled by several transducer elements placed upon an array. High-resolution digital beamforming introduces the demand for sampling rates significantly higher than the signals' Nyquist rate, which greatly increases the volume of data that must be transmitted from the system's front end. In 3D ultrasound imaging, 2D transducer arrays rather than 1D arrays are used, and more scan-lines are needed. This implies that the amount of sampled data is vastly increased with respect to 2D imaging. In this work we show that a considerable reduction in data rate can be achieved by applying the ideas of Xampling and frequency domain beamforming, leading to a sub-Nyquist sampling rate, which uses only a portion of the bandwidth of the ultrasound signals to reconstruct the image. We extend previous work on frequency domain beamforming for 2D ultrasound imaging to accommodate the geometry imposed by volumetric scanning and a 2D grid of transducer elements. We demonstrate high image quality from low-rate samples by simulation of a phantom image comprised of several small reflectors. We also apply our technique on raw data of a heart ventricle phantom obtained by a commercial 3D ultrasound system. We show that by performing 3D beamforming in the frequency domain, sub-Nyquist sampling and low processing rate are achievable, while maintaining adequate image quality.
\end{abstract}
\section{Introduction}
\label{sec:intro}
Sonography is one the most widely used imaging modalities due to its relative simplicity and radiation free operation. It uses multiple transducer elements for tissue visualization by radiating it with acoustic energy. The image is typically comprised of multiple scanlines, obtained by sequential insonification of the medium using focused acoustic beams.
Reflected signals detected at each transducer element are sampled prior to digital processing. Beamforming is the key step in image formation allowing for generation of receive sensitivity profile focused at any desired point within the image (2D) or volume (3D). The resulting beamformed signal, characterized by enhanced signal-to-noise ratio (SNR) and improved angular localization, forms a line in the image, which we refer to as beam.

\subsection{Motivation}
\label{ssec:motivation}

The aforementioned approach, used by most commercial systems today, is characterized by two important parameters --  sampling and processing rate and frame- or volume-rate.
Sampling rates required to perform high resolution digital beamforming are significantly higher than the Nyquist rate of the signal \cite{steinberg1992digital}. Taking into account the number of transducer elements and the number of lines in an image, the amount of sampled data that needs to be transferred to the processing unit and digitally processed is enormous, even in 2D imaging setups, motivating methods to reduce sampling rates.
In addition, regardless of computational power, the frame/volume-rate in this approach is limited by the time required to transmit a beam, receive and process the resulting echoes, and to repeat the process for all image lines.

Among the main focuses in the study of ultrasonic scanning is the development of real-time 3D ultrasound imaging, which overcomes major constraints of 2D imaging. 3D volume acquisition eliminates operator dependence in the imaging process -- once the 3D data set is obtained, any plane within it is available for visualization by appropriate cropping and slicing. In addition, a variety of parameters can be measured from a 3D image in a more accurate and reproducible way compared to 2D imaging \cite{prager2010three}, \cite{abo2004usefulness}, \cite{kwan2014three}. Moreover, many anatomical structures, e.g. the mitral valve, are intrinsically 3D \cite{liang2008multiplanar}, implying that their complex anatomy cannot be captured efficiently with 2D techniques.

A straight-forward approach to 3D volume acquisition is using a mechanically rotating 1D probe \cite{pini1992apparatus}. However, this technique suffers from extremely low volume rates, leading to unacceptable motion artifacts in echocardiography applications. 
Fully sampled 2D arrays, an extension of the 1D array to both lateral and elevation directions, are the most advanced technology for intrinsic 3D acquisition. Such arrays allow for significant improvement in frame rate and real-time capabilities. This is obtained by so-called 'parallel processing', namely, electronically receiving data from several points in both lateral and elevation dimensions within the 3D volume simultaneously \cite{shattuck1984explososcan, von1991high}.

Being an optimal solution in terms of frame-rate, angular resolution and SNR, fully sampled 2D arrays pose several engineering challenges \cite{prager2010three, Diarra2014study}. Due to the significantly increased number of elements, which can be as high as several thousand, the main challenge from a hardware perspective is connecting the elements to electronic channels.
In addition, the amounts of sampled data, acquired at each transmission, create a bottleneck at data transfer step and pose a severe computational burden on the digital signal processing hardware. To avoid too large connecting cables leading to unacceptable probe size and weight and to keep the electronics reasonable in power and size, as well as to reduce the data rates, numerous techniques for element number reduction have been proposed.

A straight-forward approach, referred to as sparse aperture, is to use only a subset of the 2D grid of elements upon reception and/or transmission. Several studies investigate strategies for optimal subset choices   \cite{nikolov2000application,austeng2002sparse,diarra2013design,von1991high, choe2012cmut, choe2013gpu} which limit the reduction in image quality due to energy loss and high grating and side-lobes.
In \cite{savord2003fully} Savord and Solomon present a sub-array beamforming approach allowing for significant reduction in the number of channels by sub-optimal analog beamforming, also referred to as micro-beamforming. This method was lately implemented in leading commercial systems.
Another promising method, synthetic aperture, was adopted from sonar processing and geological applications \cite{nikolov2003investigation, wygant2006beamforming}. This approach exploits multiplexing to control a fully sampled 2D array with a small number of electronic channels. Although providing improved image quality, synthetic aperture suffers from reduced frame rate and huge amounts of sampled data.

Even when reducing the number of elements the amount of sampled data is still very large due to the high number of scan-lines. Consider ultrasonic imaging of a 3D volume, using $K$ scan-lines in each one of $K$ 2D cross-sections of the volume. Scanning the entire volume yields a total of $K \times K$ scan-lines. To maintain the angular resolution in each one of the $K$ cross-sections in the 3D frame in comparison to 2D ultrasound imaging, one is forced to essentially quadrate the amount of data with respect to 2D imaging, given the same amount of transducers.

\subsection{Related work and Contributions}
\label{ssec:related work and contribution}
In this work we present an approach for data rate reduction which can be applied in conjunction with any of the existing methods for element reduction.
Our technique generalizes beamforming in frequency developed for 2D imaging.

To achieve sampling and processing rate reduction in 2D Chernyakova and Eldar \cite{chernyakova2013compressed} extended the concept of compressed beamforming \cite{wagner2012compressed} and proposed
performing beamforming in the frequency domain.
In this approach the Fourier coefficients of the beam are computed as a linear combination of those of the individual detected signals, obtained from their low-rate samples. 
When all the beam's Fourier coefficients within its bandwidth are computed, the sampling and processing rates are equal to the effective Nyquist rate. The beam in time is then obtained simply by an inverse
Fourier transform. This approach is valid without any assumptions on the ultrasound signal structure. When further rate reduction is required, only a subset of the beam's Fourier coefficients is obtained,
which is equivalent to sub-Nyquist sampling and processing.
Recovery then relies on an appropriate model of the beam, which compensates for the lack of frequency data.

The work in \cite{chernyakova2013compressed} demonstrates low-rate 2D ultrasound imaging, including the sub-Nyquist data acquisition step, low-rate processing and beamformed signal reconstruction.
Low-rate data acquisition is based on the ideas of Xampling \cite{tur2011innovation, gedalyahu2011multichannel, Baransky2012radar}, which obtains the Fourier coefficients of individual detected signals
from their low-rate samples. More specifically, using Xampling we can obtain an arbitrary and possibly nonconsecutive set $\kappa$, comprised of $K$ frequency components,
from $K$ point-wise samples of the signal filtered with an analog kernel $s^*(t)$, designed according to $\kappa$. In ultrasound imaging with modulated Gaussian pulses the transmitted signal has one
main band of energy. As a result the analog filter takes on the form of a band-pass filter, leading to a simple low-rate sampling scheme \cite{chernyakova2013compressed}. The choice of $\kappa$ dictates the bandwidth of the filter and the resulting sampling rate.

In 3D imaging the same low-rate sampling scheme can be applied to the individual signals detected at the elements of the 2D transducer, leading to considerable rate reduction, as elaborated on in
Section \ref{subsec:beam in freq reduction}. However, to benefit from the rate reduction, 3D beamforming must be performed in frequency similarly to the 2D setup.  We prove that the relationship between
the beam and the detected signals in the frequency domain, the core of beamforming in frequency, holds in the 3D imaging setup as well

In this work we derive a frequency domain formulation of beamforming that accounts for the 2D geometry of the transducer array and the 3D geometry of the medium. We show that, similarly to 2D imaging,
3D frequency domain beamforming (FDBF) can be implemented efficiently due to the decay property of the distortion function translating the dynamic beamforming time delays into the frequency domain.
When sub-Nyquist sampling and processing are applied, signal structure needs to be exploited to recover the beam from the sub-Nyquist set of its Fourier coefficients.
To this end we prove that a 3D beamformed signal obeys a finite rate of innovation (FRI) \cite{eldar2015sampling} model, just as in 2D.

We next report the results of simulations and experiments verifying the performance of the proposed method in terms of the lateral point spread functions (LPSF), axial point spread functions (APSF) and SNR. Finally we incorporate 3D FDBF to a commercial imaging system performing analog sub-array beamforming and show that the above techniques are compatible, namely, 3D FDBF does not introduce additional image degradation.

The rest of the paper is organized as follows: in Section \ref{sec:beam in time}, we review standard time-domain processing for a 3D imaging setup.
In Sections \ref{sec:beam in freq} and \ref{sec:rec} we describe the principles of 3D FDBF, image reconstruction and the achieved rate reduction. In Section \ref{sec:sim and res} the results and comparison to  time-domain beamforming are presented. Section \ref{sec:conclusion} concludes the paper.

\section{Implementation of Beamforming in Time}
\label{sec:beam in time}
Beamforming, a basic step required by all ultrasound based imaging modalities, is a common signal-processing technique that enables spatial selectivity of signal transmission or reception \cite{van2004detection}. In ultrasound imaging it allows for SNR and lateral resolution improvement.
%
Modern imaging systems transmit and receive acoustic pulses using multiple transducer elements. These elements comprise an array, generating a transmitted beam which is steered spatially by applying appropriate time delays to each element. The transducer receives acoustic pulses scattered by tissue structures, which are then sampled and processed digitally to reconstruct an image line. Reconstruction is performed with a technique known as dynamic beamforming, where the image quality is enhanced by summing the signals at individual elements after their alignment by appropriate time-delays.

\begin{figure}[htb]
\begin{minipage}[b]{1.0\linewidth}
  \centering
  \centerline{\includegraphics[width=7.5cm]{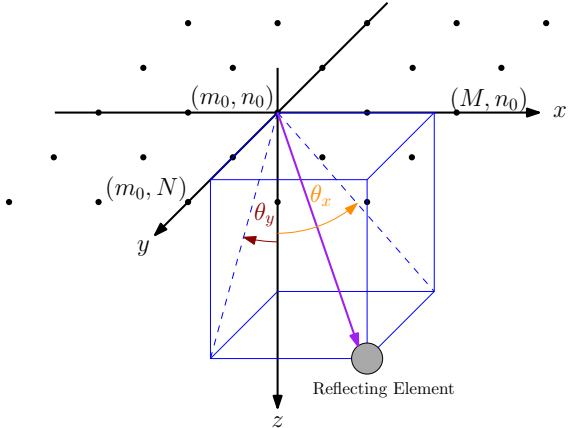}}
  \vspace{0.2cm}
\end{minipage}
\caption{$M \times N$ transducers placed in the $x$-$y$ plane. An acoustic pulse is transmitted in a direction $\theta_x, \theta_y$. The echoes scattered from perturbations in the radiated tissue are received by the array elements.}
\label{fig:array}
\end{figure}

To derive frequency-domain implementation of 3D beamforming we begin by introducing standard time-domain processing.
Consider a grid of $M \times N$ transducers located in the $x$-$y$ plane, depicted in Fig. \ref{fig:array}. The geometry imposed by 3D ultrasound imaging requires the use of two steering angles
and thus a 2D array of transducers. The entire grid transmits pulses into the tissue.  We note that the grid may have a small curvature along the $z$ axis, so the array elements do not lie in the same plane. For the sake of simplicity, this type of curvature is not displayed in Fig. \ref{fig:array}.
%

We choose a reference element, $(m_0,n_0)$, placed at the origin, and denote the distances along the $x$ and $y$ axes to the $(m,n)$ element by $\delta_m,\delta_n$, respectively. We also denote the height
of the $(m,n)$ element with respect to the origin by $\delta_{m,n}^z$ for the case where there exists a curvature along the $z$ axis. Note that we assume $\delta_{m_0,n_0}^z=0$, so that the reference element
is not necessarily included in the transducer grid, and is defined for mathematical convenience. Let us consider a pulse transmitted along a scan-line specified by spatial angles $\theta_x,\theta_y$.
Setting $t=0$ at the moment of transmission from the $(m_0,n_0)$ element, it can be shown that at time $t \ge 0$ the pulse reaches the coordinates:
\begin{equation} \label{coordinates 1}
(x(t),y(t),z(t)) = ct(x_{\theta}, y_{\theta}, z_{\theta}),
\end{equation}
with
\begin{align}
\label{coordinates 2}
\vspace{-0.1cm}
x_{\theta} &= \frac{\sin{\theta_x}\cos{\theta_y}}{\sqrt{1-\sin^2{\theta_x}\sin^2{\theta_y}}} \nonumber \\
y_{\theta} &= \frac{\cos{\theta_x}\sin{\theta_y}}{\sqrt{1-\sin^2{\theta_x}\sin^2{\theta_y}}} \\
z_{\theta} &= \frac{\cos{\theta_x}\cos{\theta_y}}{\sqrt{1-\sin^2{\theta_x}\sin^2{\theta_y}}}. \nonumber
\vspace{-0.1cm}
\end{align}
Here $c$ is the propagation velocity in the medium. A point reflector located at this position scatters the energy, such that the echo is detected by all array elements at a time depending on their locations.

Denote by $\varphi_{m,n}(t;\theta_x,\theta_y)$ the signal detected by the $(m,n)$ element and by $\hat{\tau}_{m,n}(t;\theta_x,\theta_y)$ the time of detection. Then:
\begin{equation}
\label{tau^m}
\hat{\tau}_{m,n}(t;\theta_x,\theta_y)=t+\frac{d_{m,n}(t;\theta_x,\theta_y)}{c},
\end{equation}
%
where
\begin{align}
\label{d mn}
d_{m,n}(t;\theta_x,\theta_y)= & \\*
 &\hspace{-1.0cm}  \sqrt{(x(t)-\delta_m)^2 + (y(t)-\delta_n)^2 + (z(t)-\delta_{m,n}^z)^2} \nonumber
\end{align}
is the distance traveled by the reflection.
Beamforming involves summing the signals detected by multiple receivers while compensating for the differences in detection time.

Using \eqref{tau^m}, the detection time at $(m_0,n_0)$ is $\hat{\tau}_{m_0,n_0}(t;\theta_x,\theta_y)=2t$ since $\delta_{m_0}=\delta_{n_0}=\delta_{m,n}^z=0$. We wish to apply a delay to $\varphi_{m,n}(t;\theta_x,\theta_y)$ such that the resulting signal, denoted by $\hat{\varphi}_{m,n}(t;\theta_x,\theta_y)$, satisfies:
\begin{equation*}
\hat{\varphi}_{m,n}(2t;\theta_x,\theta_y)=\varphi_{m,n}(\hat{\tau}_{m,n}(t;\theta_x,\theta_y);\theta_x,\theta_y).
\end{equation*}
Doing so, we can align the reflection detected by the $(m,n)$ receiver with the one detected at $(m_0,n_0)$. Denoting $\tau_{m,n}(t;\theta_x,\theta_y)$ $=$ $\hat{\tau}_{m,n}(t/2;\theta_x,\theta_y)$  and using \eqref{tau^m}, the following aligned signal is obtained:
\vspace{-0.1cm}
%
\begin{align}
\label{phim}
\hat{\varphi}_{m,n}(t;\theta_x,\theta_y)&=\varphi_{m,n}(\tau_{m,n}(t;\theta_x,\theta_y);\theta_x,\theta_y), \nonumber \\
\tau_{m,n}(t;\theta_x,\theta_y)&= \\* \nonumber
& \hspace{-1.5cm} \frac{1}{2}\left(t+\sqrt{t^2 + 4|\gamma_{m,n}|^2 -4t\left( \gamma_m x_{\theta} + \gamma_n y_{\theta} + \gamma_{m,n}^z z_{\theta} \right) }\right),
\vspace{-0.1cm}
\end{align}
%
where we defined $\gamma_m=\delta_m/c$, $\gamma_n=\delta_n/c$, $\gamma_{m,n}^z=\delta_{m,n}^z/c$ and $|\gamma_{m,n}|=\sqrt{\gamma^2_m+\gamma^2_n+(\gamma_{m,n}^z)^2}$.

The beamformed signal may now be derived by averaging the aligned signals. We assume that the echo reception process involves a subset of the transducer array, denoted by
$\mathcal{M} \subseteq \{(m,n) | ~1 \le m \le M, 1 \le n \le N\}$:
\begin{equation}
\label{phi beamformed}
\vspace{-0.1cm}
\Phi(t;\theta_x,\theta_y)=\frac{1}{N_{\textrm{RX}}}\sum_{(m,n) \in \mathcal{M}}{\hat{\varphi}_{m,n}(t;\theta_x,\theta_y)}.
\vspace{-0.1cm}
\end{equation}
Here ${N_{\textrm{RX}}}$$=$$|\mathcal{M}|$ is the number of transducers participating in the reception process. We note that in order to obtain optimal performance in terms of SNR and angular resolution,
all transducer elements should be used. However, as mentioned in Section \ref{sec:intro}, the number of active elements is often reduced due to practical constraints.

The beamforming process is carried out digitally, rather than by manipulation of the analog signals. The signals detected at each element must be sampled at a sufficiently high rate to apply high-resolution
time shifts defined in \eqref{phim}. This implies that the signal is sampled at rates significantly higher than its Nyquist rate, in order to improve the system's beamforming resolution and to avoid artifacts caused by digital implementation of beamforming in time. From now on we will denote this rate as the beamforming rate $f_s$, which usually varies from 4 to 10 times the transducer central frequency \cite{steinberg1992digital,chernyakova2013compressed}.


We conclude this section by evaluating the number of samples typically required to obtain a single volume for some predefined image depth. Our evaluation is based on the imaging setup used in the simulation 
displayed in Section \ref{ssec:simulation setup}. The simulation assumes an ultrasonic scanner comprising a 32$\times$32 grid of transducers, all of which are active both on transmission and reception ($N_{\textrm{RX}}=1024$). Such an array constitutes a reference for comparison of image quality resulting from different methods for data rate reduction. The radial depth of the scan is set as $r=5.5$ cm with a speed of sound of $c = 1540 \textrm{}m/\textrm{sec}$, yielding a time of flight of $T=2r/c\simeq71.43 ~\mu$sec. The acquired signal is characterized by a band-pass bandwidth of 1.4 MHz centered at a carrier frequency of $f_{0}=3$ MHz. It is sampled at a rate of $f_s=18.25$~MHz to provide a sufficient beamforming resolution leading to $N=1304$ samples taken at each transducer.
Every frame contains $21 \times 21$ scan-lines, such that the scanned volume is a square pyramid with an opening angle of $14.3^{\circ}$. This set of scanning angles is a relatively narrow set with a typical margin between subsequent beam lines.
Therefore, assuming that it is possible to sample all 1024 elements to obtain optimal image quality, the total number of samples that must be processed to display a single frame is
$21 \times 21 \times 1024 \times 1304 = 5.89 \cdot 10^8$.
This number of samples is huge even for a moderate imaging depth of $5.5$ cm; the imaging depth typically required for cardiac imaging is around $16$ cm. Achieving a reasonable frame-rate using such an amount of samples is infeasible for any low-cost ultrasound machine.
Therefore, even assuming a hardware solution allowing for connection of all the transducer elements to electronic channels, the amount of data is still a bottleneck.
As a result sparsely populated arrays of transducer elements may be used. This typically causes a reduction in angular resolution and, more significantly, low SNR.
A solution that reduces the amount of samples while using the entire transducer grid in the reception stage will address this problem.

\section{Beamforming in Frequency}
\label{sec:beam in freq}

To substantially reduce the number of samples taken at each transducer element we aim to use the low-rate sampling scheme proposed in \cite{chernyakova2013compressed}.
To this end we derive a frequency-domain formulation of 3D beamforming allowing to compute the Fourier coefficients of the beam from the detected signals' low-rate samples.
In this section we show that similarly to 2D imaging the Fourier coefficients of the 3D beam can be computed as a linear combination of the Fourier confucianists of the received signals. We note that due to the dynamic nature of beamforming, such a relationship is not trivial and requires appropriate justification.

\subsection{Beamforming in Frequency Scheme}
\label{subsec:beam in freq scheme}


We start from the computation of the Fourier series coefficients of the beamformed signal $\Phi(t;\theta_x,\theta_y)$.
It is shown in Appendix \ref{sec:app beamformed support} that the support of $\Phi(t;\theta_x,\theta_y)$ is limited to $\left[0,T_B(\theta_x,\theta_y)\right)$, where $T_B(\theta_x,\theta_y)$ is given by:
\begin{equation} \label{TB}
T_B(\theta_x,\theta_y) = \underset{(m,n) \in \mathcal{M}}{\textrm{min}} \tau_{m,n}^{-1}(T;\theta_x,\theta_y),
\end{equation}
with $\tau_{m,n}(t;\theta_x,\theta_y)$ defined in \eqref{phim}. It is also shown that $T_B(\theta_x,\theta_y) \le T$, where $T$ is defined by the transmitted pulse penetration depth.

Consider the Fourier series of the beamformed signal, $\{c[k]\}_k$, in the interval $[0,T]$:
\begin{equation}
\label{fourier coeff of beamformed 1}
c[k]=\frac{1}{T}\int_0^T \Phi(t;\theta_x,\theta_y) I_{[0,T_B(\theta_x,\theta_y))} \expk dt,
\end{equation}
where $I_{[a,b]}$ is the indicator function, plugged in to cancel noise since the useful information in $\Phi(t;\theta_x,\theta_y)$ is restricted to $\left[0,T_B(\theta_x,\theta_y)\right)$.
In order to find a relation between $c[k]$ and the Fourier coefficients of $\varphi_{m,n}(t;\theta_x,\theta_y)$, we substitute \eqref{phi beamformed} into \eqref{fourier coeff of beamformed 1}:
\begin{align}
\label{fourier coeff of beamformed 2}
c[k] &= \frac{1}{N_{\textrm{RX}}} \sum_{(m,n) \in \mathcal{M}} \frac{1}{T}\int_0^T {\hat{\varphi}_{m,n}(t;\theta_x,\theta_y)} I_{[0,T_B(\theta_x,\theta_y))} \expk dt \nonumber \\*
&=\frac{1}{N_{\textrm{RX}}} \sum_{(m,n) \in \mathcal{M}}\hat{c}_{m,n}[k],
\end{align}
where we defined
\begin{align}
\label{c k m def}
\vspace{-0.1cm}
\hat{c}_{m,n}[k]&=\frac{1}{T}\int_0^T {\varphi_{m,n}(\tau_{m,n}(u;\theta_x,\theta_y);\theta_x,\theta_y)} \nonumber \\*
&\hspace{0.8cm} \times I_{[0,T_B(\theta_x,\theta_y))} e^{-i \frac{2\pi}{T}k u} du.
\vspace{-0.1cm}
\end{align}
Substituting the integration variable $u$ with $\tau=\tau_{m,n}(u;\theta_x,\theta_y)$ we get
\begin{align*}
\label{t tau sub}
\vspace{-0.1cm}
u&=\frac{\tau^{2}-\left|\gamma_{m,n}\right|^{2}}{\tau-\left(\gamma_{m}x_{\theta}+\gamma_{n}y_{\theta}+\gamma_{m,n}^{z}z_{\theta}\right)}, \\*
du&=\dfrac{\tau^{2}+\left|\gamma_{m,n}\right|^{2}-2\tau\cdot\left(\gamma_{m}x_{\theta}+\gamma_{n}y_{\theta}+\gamma_{m,n}^{z}z_{\theta}\right)}{\left[\tau-\left(\gamma_{m}x_{\theta}+\gamma_{n}y_{\theta}+\gamma_{m,n}^{z}z_{\theta}\right)\right]^{2}}d\tau,
\vspace{-0.1cm}
\end{align*}
where $x_{\theta},y_{\theta},z_{\theta}$ are defined in \eqref{coordinates 2}.
Plugging this into \eqref{c k m def} and renaming the integration variable $\tau \rightarrow t$, result in
\begin{equation}
\label{c k m}
\vspace{-0.2cm}
\hat{c}_{m,n}[k]=\frac{1}{T}\int_0^T q_{k,m,n}(t;\theta_x,\theta_y) \varphi_{m,n}(t;\theta_x,\theta_y) \expk  dt
\vspace{-0.1cm}
\end{equation}
with
\begin{flalign}
\label{g j m q j m}
q_{k,m,n}(t;\theta_x,\theta_y)&= I_{[|\gamma_{m,n}|,\tau_{m,n}(T_B(\theta_x,\theta_y);\theta_x,\theta_y))}(t) \times \nonumber \\*
&\hspace{-2.2cm}\frac{t^2+|\gamma_{m,n}|^2-2t\cdot\left( \gamma_m x_{\theta} + \gamma_n y_{\theta} + \gamma_{m,n}^z z_{\theta} \right)}{\left(t-\left( \gamma_m x_{\theta} + \gamma_n y_{\theta} + \gamma_{m,n}^z z_{\theta} \right)\right)^2} \times \\*
&\hspace{-2.2cm}\exp\left\{-i\frac{2\pi}{T}k\left(\frac{t\cdot\left( \gamma_m x_{\theta} + \gamma_n y_{\theta} + \gamma_{m,n}^z z_{\theta} \right)-|\gamma_{m,n}|^2}{t-\left( \gamma_m x_{\theta} + \gamma_n y_{\theta} + \gamma_{m,n}^z z_{\theta} \right)}\right)\right\}. \nonumber
\end{flalign}

Note that in contrast to \eqref{c k m def}, \eqref{c k m} contains a non-delayed version of $\varphi_{m,n}(t;\theta_x,\theta_y)$, while the delays are applied through the distortion function
$q_{k,m,n}(t;\theta_x,\theta_y)$, defined in \eqref{g j m q j m}.
This allows us to express $\varphi_{m,n}(t;\theta_x,\theta_y)$ in terms of its Fourier series coefficients, denoted by $c_{m,n}[l]$. We also make use of the Fourier coefficients of $q_{k,m,n}(t;\theta_x,\theta_y)$ with respect to $[0,T]$, denoted by $Q_{k,m,n;\theta_x,\theta_y}[l]$, and rewrite \eqref{c k m} as follows:
\begin{align}\label{c k m fourier}
\vspace{-0.1cm}
 \hat{c}_{m,n}[k]&=\sum_l c_{m,n}[l]\frac{1}{T}\int_0^T q_{k,m,n}(t;\theta_x,\theta_y) e^{-i\frac{2\pi}{T}(k-l)}  dt \\ \nonumber
 &=\sum_l c_{m,n}[k-l]Q_{k,m,n;\theta_x,\theta_y}[l].
 \vspace{-0.1cm}
\end{align}
The substitution of the distortion function by its Fourier coefficients effectively transfers the beamforming delays defined in \eqref{phim} to the frequency domain. We note that $q_{k,m,n}(t;\theta_x,\theta_y)$ is independent of the received signals, namely, it is defined solely by the array geometry. Its Fourier coefficients, therefore, are computed off-line and stored as a look-up-table (LUT).

According to Proposition 1 in \cite{wagner2012compressed}, which can be easily extended to the 3D imaging setup, $\hat{c}_{m,n}[k]$ can be approximated sufficiently well when we replace the infinite summation in \eqref{c k m fourier} by a finite one:
\begin{equation}\label{c k m approx explicit}
\vspace{-0.1cm}
\hat{c}_{m,n}[k]\simeq\sum_{l = -L_1}^{L_2}c_{m,n}[k-l]Q_{k,m,n;\theta_x,\theta_y}[l].
\vspace{-0.1cm}
\end{equation}
The Fourier coefficients of the beam, $c[k]$, can now be easily calculated by plugging \eqref{c k m approx explicit} into \eqref{fourier coeff of beamformed 2}:
\begin{equation}
\label{fourier coeff of beamformed 3}
c[k] \simeq \frac{1}{N_{\textrm{RX}}} \sum_{(m,n) \in \mathcal{M}}\sum_{l = -L_1}^{L_2}c_{m,n}[k-l]Q_{k,m,n;\theta_x,\theta_y}[l].
\end{equation}

\begin{figure}[htb]
\begin{minipage}[b]{1.0\linewidth}
  \centering
  \centerline{\includegraphics[width=8.5cm]{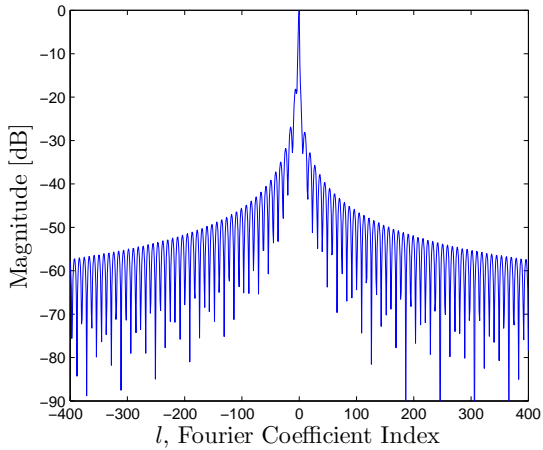}}
\end{minipage}
\caption{$\{ Q_{k,m,n;\theta_x,\theta_y}[l] \}$, the Fourier coefficients of $q_{k,m,n}(t;\theta_x,\theta_y)$, display a rapid decay rapidly around the DC component.}
\label{fig:Q}
\end{figure}

The approximation in \eqref{c k m approx explicit} relies on the decay properties of $\{ Q_{k,m,n;\theta_x,\theta_y}[l] \}$.
According to the results reported in \cite{chernyakova2013compressed} most of the energy of the Fourier coefficients of the 2D distortion function is concentrated around the DC component, allowing for
efficient implementation of beamforming in frequency. This decaying property is retained in 3D beamforming: numerical studies show that most of the energy of $\{ Q_{k,m,n;\theta_x,\theta_y}[l] \}$ is
concentrated around the DC component, irrespective of the choice of $k,m,n,\theta_x,\theta_y$. We assume that for $l<-L_1$ and $l>L_2$, $\{ Q_{k,m,n;\theta_x,\theta_y}[l] \}$ are several orders of magnitude lower and thus can be neglected. The choice of $L_1,L_2$ controls the approximation quality. We display these decay properties in Fig. \ref{fig:Q}, where $Q_{k,m,n;\theta_x,\theta_y}[l]$ is plotted as a function of $l$ for $k=300$, $m=7$, $n=21$, $\theta_x=0.28~[\textrm{rad}]$ and $\theta_y=0.36~[\textrm{rad}]$. 

\subsection{Rate Reduction by Beamforming in Frequency}
\label{subsec:beam in freq reduction}
We now show that FDBF allows one to generate a frame using a reduced number of samples of the individual signals with respect to time-domain beamforming. When the signal's structure is not considered,
this is done by avoiding the oversampling factor required by digital implementation of time-domain beamforming. In this case the processing is performed at the effective Nyquist rate defined by the signal's
effective bandwidth. Further rate reduction can be obtained when the FRI structure of the beamformed signal is taken into account and compressed sensing (CS) techniques are used for
its recovery \cite{eldar2012compressed,eldar2015sampling}.

As can be seen in \eqref{fourier coeff of beamformed 3}, in order to calculate an arbitrary set $\kappa$ of size $\mathcal{K}$ of Fourier coefficients of the beamformed signal, only $\mathcal{K}+L_1+L_2$ Fourier coefficients of each one of the individual signals are required. The image line is then reconstructed from the beamformed signal's Fourier coefficients $\left\{ c[k] \right\}$.
Calculating the entire set of Fourier coefficients in the bandwidth of the beamformed signal $\beta$, $|\beta|=B$, implies $B\gg L_1+L_2$ and, therefore, allows one to
obtain all the information in the frequency domain while avoiding oversampling required by time-domain beamforming. This is due to the fact that the low-rate sampling scheme described in Section \ref{ssec:related work and contribution} obtains $B+L_1+L_2\approx B$ Fourier coefficients of the individual signals required for FDBF from their $B$ low-rate samples. 
%
%
%
Thus, performing FDBF by calculating the entire bandwidth of the beamformed signal achieves an approximately $N/B$ rate reduction factor with respect to time-domain beamforming, where $B$ is the number of Fourier coefficients in the bandwidth of the beamformed signal and $N$ is the number of samples required by the beamforming rate $f_s$.

Further rate reduction is possible by acquiring a part of the bandwidth of the beamformed signal, $\mu \subset \beta,~|\mu|=M$. We may calculate it from $M+L_1+L_2\approx M$ samples of the individual signals, which are sampled at a rate that is $N/M$ lower than the standard beamforming rate $f_s$.
In Section \ref{sec:rec}, we take advantage of the beamformed signal structure to reconstruct the beam from its partial frequency data. A detailed discussion on the achieved rate reduction is provided in Section \ref{ssec:simulation setup}.

\vspace{-0.1cm}
\section{Recovery Method from Sub-Nyquist Samples}
\label{sec:rec}
When all the beam's Fourier coefficients within its effective bandwidth are computed, the beam in time is recovered by an inverse Fourier transform. When only a subset of the coefficients is obtained by sub-Nyquist sampling and processing, we exploit the structure of the beam to reconstruct it from its partial frequency data.

According to \cite{tur2011innovation}, we can model the detected signals at the individual transducer elements, $\{ \varphi_{m,n}(t;\theta_x,\theta_y) \}_{(m,n) \in \mathcal{M}}$, as FRI signals. That is, we assume that the individual signals can be regarded as a sum of pulses, all replicas of a known transmitted pulse shape:
\begin{equation}\label{individual FRI}
\vspace{-0.1cm}
\varphi_{m,n}(t;\theta_x,\theta_y) = \sum_{l=1}^L \tilde{a}_{l,m,n} h(t-t_{l,m,n}).
\vspace{-0.1cm}
\end{equation}
Here $h(t)$ is the transmitted pulse shape, $L$ is the number of scattering elements in the direction of the transmitted pulse $(\theta_x,\theta_y)$, $\{\tilde{a}_{l,m,n}\}_{l=1}^L$ are the unknown amplitudes of the reflections and $\{t_{l,m,n}\}_{l=1}^L$ are the times at which the reflection from the $l$-th scatterer arrives at the $(m,n)$ element.

It is shown in Appendix \ref{sec:app beamformed FRI} that the beamformed signal in 3D imaging approximately satisfies the FRI model, just as it does in 2D imaging \cite{wagner2012compressed}. Namely, it can be written as
\begin{equation}\label{beam FRI}
\vspace{-0.1cm}
\Phi(t;\theta_x,\theta_y)\simeq\sum_{l=1}^L \tilde{b}_l h(t-t_l),
\vspace{-0.1cm}
\end{equation}
where $h(t)$ and $L$ are defined as above, $\{\tilde{b}_l\}_{l=1}^L$ are the unknown amplitudes of the reflections and $\{t_l\}_{l=1}^L$ are the times at which the reflection from the $l$-th scatterer arrives at the reference element $(m_0,n_0)$.

Having acquired the Fourier coefficients $c[k]$ as described in the previous section, we now wish to reconstruct the beamformed signal. Since the beam satisfies the FRI model our task is to extract the unknown parameters, $\{\tilde{b}_l\}_{l=1}^L$ and $\{t_l\}_{l=1}^L$ that completely describe it.

Using \eqref{beam FRI} the Fourier coefficients of $\Phi(t;\theta_x,\theta_y)$ are given by:
\begin{align}
\label{Phi Fourier coeff}
c[k] &= \frac{1}{T}\int_{0}^{T} \Phi(t;\theta_x,\theta_y) e^{-i\frac{2\pi}{T}kt} \nonumber \\*
&\simeq \frac{1}{T}\int_{0}^{T} \left( \sum_{l = 1}^{L} \tilde{b}_l h(t-t_l) \right) e^{-i\frac{2\pi}{T}kt} \nonumber \\*
&= \sum_{l = 1}^{L} \tilde{b}_l \left( \frac{1}{T}\int_{0}^{T} h(t-t_l) e^{-i\frac{2\pi}{T}k(t-t_l)} \right) e^{-i\frac{2\pi}{T}kt_l} \nonumber \\*
&= h[k] \sum_{l = 1}^{L} \tilde{b}_l e^{-i\frac{2\pi}{T}kt_l},
\end{align}
where $h[k]$ is the $k$-th Fourier coefficient of $h(t)$.

By quantizing the delays $\{t_l\}_{l=1}^L$ with quantization step $T_s = \frac{1}{f_s}$, such that $t_l = q_l T_s$ for $q_l \in \mathbb{Z}$, we may write the Fourier coefficients of the beamformed signal as:
\begin{equation}\label{Phi Fourier coeff quant}
\vspace{-0.1cm}
c[k] = h[k] \sum_{l=0}^{N-1} b_l e^{-i\frac{2\pi}{N}kl},
\vspace{-0.1cm}
\end{equation}
with $N=\lfloor T/T_s\rfloor$, $b_l=\tilde{b}_l\delta_{l,q_l}$ and $\delta_{a,b}$ is the Kronecker delta. We conclude that recovering the beamformed signal in time is equivalent to determining $b_l$ in \eqref{Phi Fourier coeff quant} for $0\le l \le N-1$.
In vector-matrix notation \eqref{Phi Fourier coeff} cab be rewritten as:
\begin{equation}\label{matrix}
\vspace{-0.1cm}
\mathbf{c}=\mathbf{HDb}=\mathbf{Ab},
\vspace{-0.1cm}
\end{equation}
where $\mathbf{c}$ is a vector of length $\mathcal{K}$ with $k$-th entry $c[k]$, $\mathbf{H}$ is a $\mathcal{K}\times\mathcal{K}$ diagonal matrix with $k$-th entry $h[k]$
, $\mathbf{D}$ is a $\mathcal{K}\times N$ matrix whose rows are taken from the $N\times N$ DFT matrix corresponding to the relevant Fourier indices of $\Phi(t;\theta_x,\theta_y)$, and $\mathbf{b}$ is a column vector of length $N$ with $l$-th entry $b_l$.

We wish to extract the values of $\mathbf{b}$, which fully describe the beamformed signal. To do so, we rely on the assumption that a typical ultrasound image is comprised of a relatively small number of strong reflectors in the scanned tissue. In other words, we assume the vector $\mathbf{b}$ to be compressible, similarly to \cite{chernyakova2013compressed}. We then find $\mathbf{b}$ by solving an $\ell_1$ optimization problem:
\begin{equation}\label{l1}
\vspace{-0.1cm}
\underset{\mathbf{b}}{\min}\|\mathbf{b}\|_{1}\mbox{ s.t.}\mbox{ }\|\mathbf{Ab-c}\|_{2}\le\varepsilon.
\vspace{-0.1cm}
\end{equation}
In practice, we solve \eqref{l1} using the NESTA algorithm \cite{becker2011nesta} which works well when the signal of interest has high dynamic range. NESTA uses a single smoothing parameter, $\mu$, selected based on a trade-off between accuracy and speed of convergence. We choose this parameter empirically to achieve optimal performance with respect to image quality.

To summarize this section, a step-by-step description of the 3D low-rate imaging process is given in Algorithm \ref{alg:rec alg}.

\begin{algorithm}
\caption{Image acquisition algorithm}
\label{alg:rec alg}
\begin{algorithmic}[1]

    \STATE Calculate the Fourier coefficients of $q_{j,m,n}\left(t;\theta_{x},\theta_{y}\right)$, defined in \eqref{g j m q j m}. This calculation is performed offline and does not affect the system's real-time performance.
    \STATE Choose the approximation quality by determining $L_{1},L_{2}$, defined according to the decay properties of $\left\{Q_{k,m,n;\theta_{x},\theta_{y}}\left[l\right]\right\} _{l}$, displayed in Fig. \ref{fig:Q}. An adequate approximation can be performed by choosing $L_{1},L_{2}$ to be no greater than 10.
    \STATE Choose a subset $\kappa$ of Fourier coefficients of the beamformed signal to be used in reconstruction.
    \STATE Acquire the Fourier coefficients of the individual signals relevant for reconstruction from low-rate samples, according to \cite{chernyakova2013compressed}. At each transducer element $(m,n)\in\mathcal{M}$, $\left\{ c_{m,n}\left[l\right]\right\} _{l=k_{1}+L_{1}}^{k_{2}+L_{2}}$,
where $k_{1},k_{2}$ are the lowest and highest indices in the subset.
    \STATE Perform the calculation in \eqref{c k m approx explicit}.
    \STATE Compute the beamformed signal's Fourier series coefficients:
    \[
    c[k]=\dfrac{1}{N_{\textrm{RX}}}\sum_{\left(m,n\right)\in\mathcal{M}}\hat{c}_{m,n}[k].
    \]
    \\
    \STATE Solve the optimization problem \eqref{l1} to
extract the vector $\mathbf{b}$ that characterizes the beamformed signal.
    \STATE Incorporate the known temporal shape of the pulses, $h\left(t\right)$,
onto the vector $\mathbf{b}$, and perform standard postprocessing steps, such as log-compression and interpolation.

\end{algorithmic}
\end{algorithm}

\vspace{-0.1cm}
\section{Simulations and Results}
\label{sec:sim and res}

To analyze the performance of the outlined methodology relative to standard time-domain beamforming in a manner independent of the specifics of any individual system, a \emph{k-Wave} \cite{treeby2010kwave} simulation of a 3D ultrasound system is presented.
We first simulate the acoustic imaging of a noise free volume containing three point scatterers and analyze the effect of the achieved rate reduction on lateral and axial point spread functions. The performance of low-rate 3D FDBF is compared to that of standard time-domain processing. As mentioned in Section \ref{sec:intro}, the number of samples can be reduced when only a partial set of transducer elements is used upon reception. To compare this approach to the proposed method, time-domain beamforming is also performed using the data collected only from the elements placed along the array's main diagonals.
We next show the advantage of 3D FDBF in the presence of noise compared to time-domain beamforming with the reduced number of elements.
Finally we show that the proposed approach can be incorporated into a commercial imaging system performing sub-array beamforming to reduce the number of channels. The results verify that rate reduction obtained by 3D FDBF does not introduce further image quality degradation.

\subsection{Simulation Setup}
\label{ssec:simulation setup}
We simulate acoustic imaging of a volume of size $28~\textrm{mm} \times 28~\textrm{mm} \times 55~\textrm{mm}$. The volume contains three point reflectors, placed at depths $26$, $31.5$ and $37$ mm from the center of a square planar 2D transducer grid. The reflectors are located around $\theta_x = -7.5^{\circ}$,  $0^{\circ}$ and $7.5^{\circ}$, respectively, with $\theta_y  = 0^{\circ}$. The reflector at depth $31.5$ mm is located at the focus point of the transmitted pulse.
The array is comprised of $32\times32 = 1024$ elements, spaced $140$ $\mu$m apart. The central pulse frequency is $3$ MHz with a bandwidth of $1.4$ MHz and the sampling rate is $f_s=18.25$ MHz. A penetration depth of $T=2r/c\simeq71.43$ $\mu$sec yields $N=1304$ samples at each transducer element, so that the bandwidth of the beamformed signal contains $B=200$ Fourier coefficients. A single volume comprises $21\times21$ scanned angles.

Denote by $\kappa$ of cardinality $\mathcal{K}$ the set of Fourier coefficients of the beam, obtained by the proposed method. To verify the performance for different rate reduction factors the collected data is processed using our technique with $\mathcal{K} = B$, $\mathcal{K} = B/2$ and $\mathcal{K} = B/3$ corresponding to the entire, half and one third of the effective bandwidth respectively. The results are compared to those obtained by time-domain beamforming performed both using full and diagonal grids upon reception.

First, we assess the amount of samples required to obtain a volume. For time-domain beamforming using the full grid, we must process $21\times21\times1024\times N=5.89\cdot10^9$ samples. Using only the main diagonals of the transducer grid, this amount reduces to $21\times21\times64\times N=36.81\cdot10^6$ samples.
Applying FDBF, the reconstruction relied on $\mathcal{K} = 200$, $\mathcal{K} =~100$ and $\mathcal{K} = 67$ Fourier coefficients of the beamformed signal. To calculate these coefficients, as described in Section \ref{sec:beam in freq}, we chose $L_1=L_2=10$. The  total amount of Fourier coefficients required at each transducer is $\nu = \mathcal{K} + L_1 + L_2$. A mechanism proposed in \cite{chernyakova2013compressed} allows us to obtain these coefficients from $\nu$ samples of individual detected signals. Thus, a single volume can be produced by processing a total of $21 \times 21 \times 1024 \times \nu$ samples. The total number of samples required for each processing method is displayed in Table \ref{tab:num of samples}. We note that FDBF, using about half the bandwidth of the beam, is comparable to time-domain processing using only the diagonal elements of the grid in terms of processing rate.

\begin{figure*}[htb]
    \begin{minipage}[b]{.2\linewidth}
        \centering \includegraphics[width = 5.5cm, angle = 270]{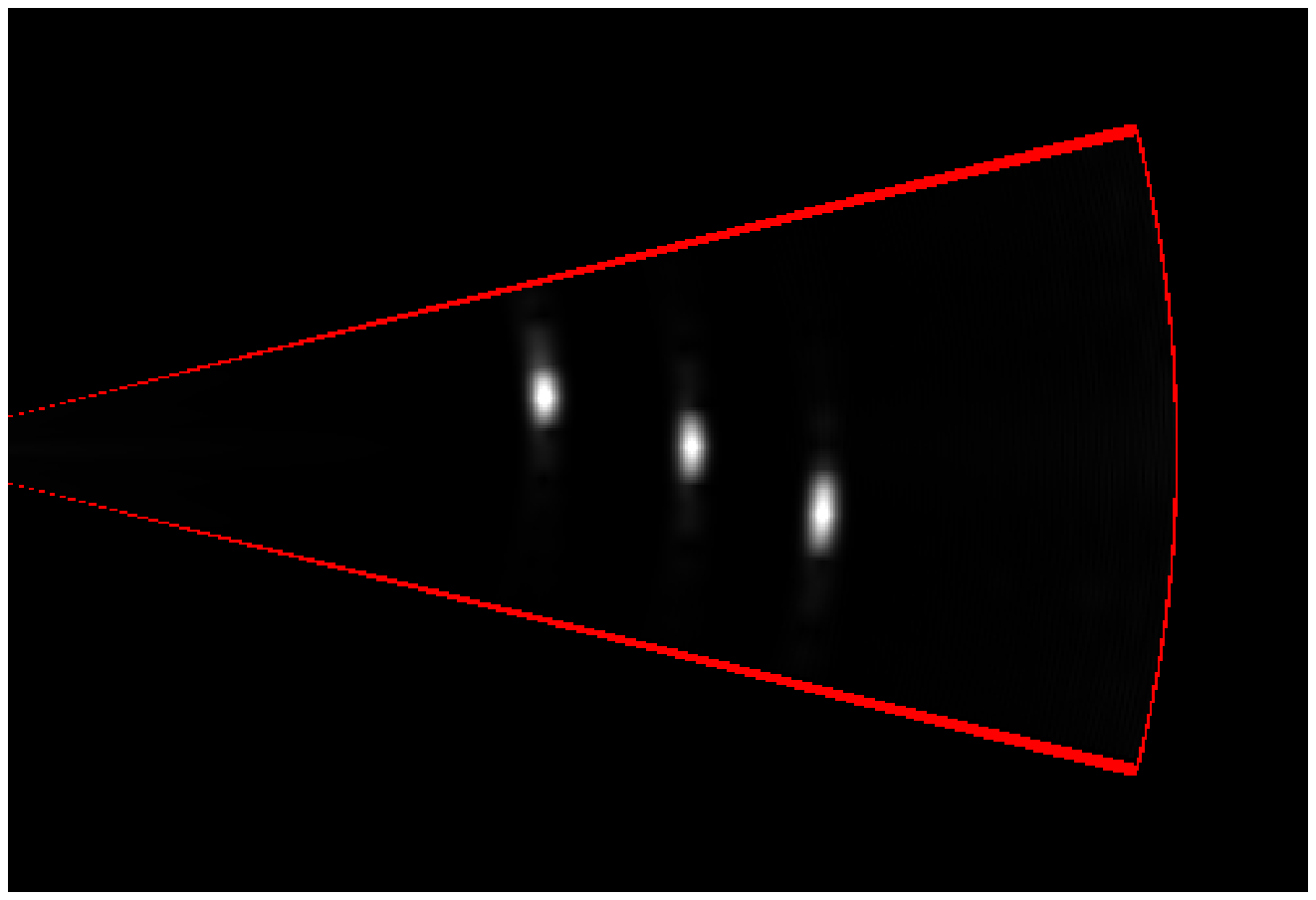}
        \vspace{-0.7cm}
        \subcaption{}
        \label{subfig:kWave Y=0 time full}
    \end{minipage}
    \hspace{-0.4cm}
    \hfill
    \begin{minipage}[b]{.2\linewidth}
        \centering \includegraphics[width = 5.5cm, angle = 270]{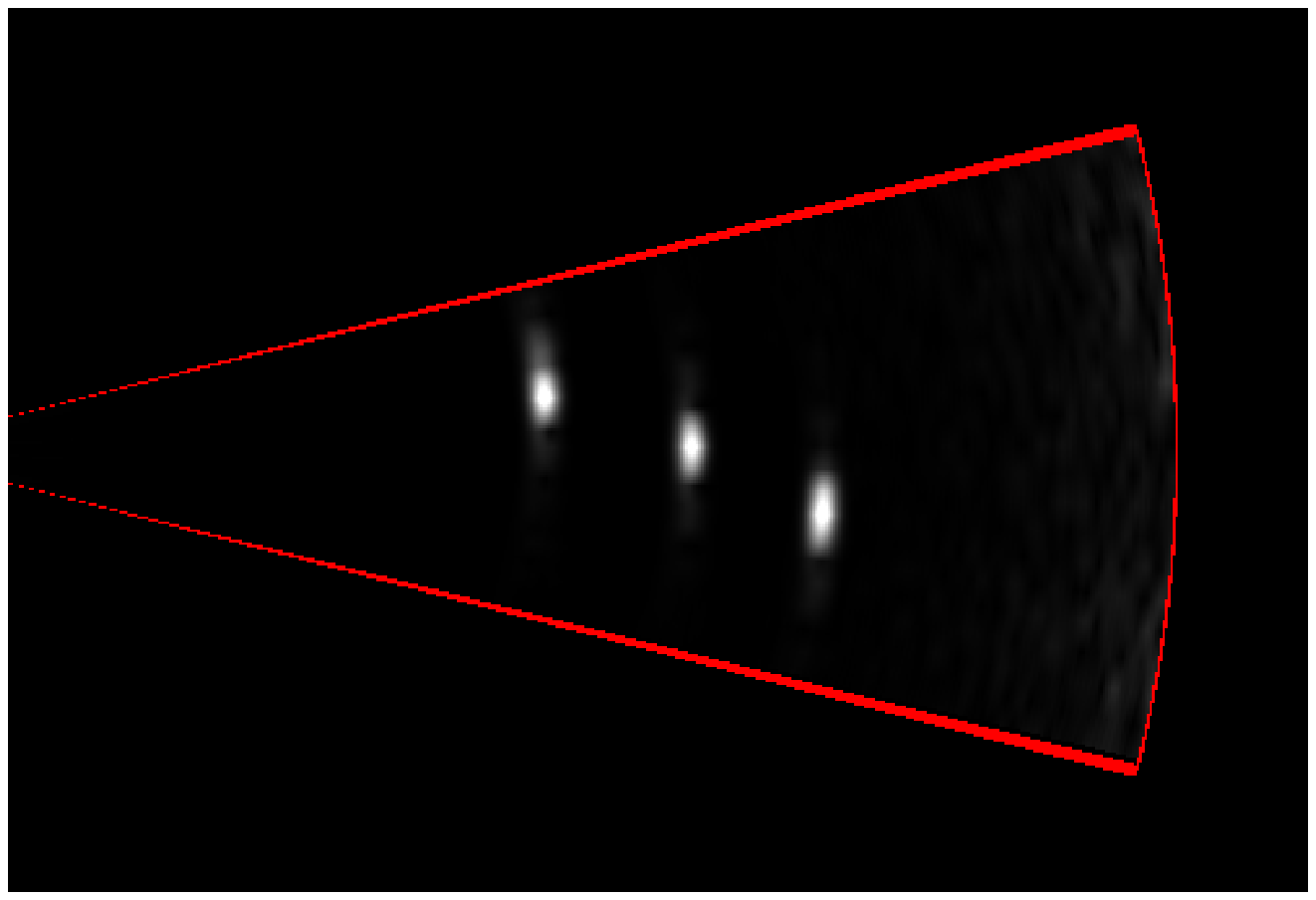}
        \vspace{-0.7cm}
        \subcaption{}
        \label{subfig:kWave Y=0 time diagonals}
    \end{minipage}
    \hspace{-0.4cm}
    \hfill
    \begin{minipage}[b]{.2\linewidth}
        \centering \includegraphics[width = 5.5cm, angle = 270]{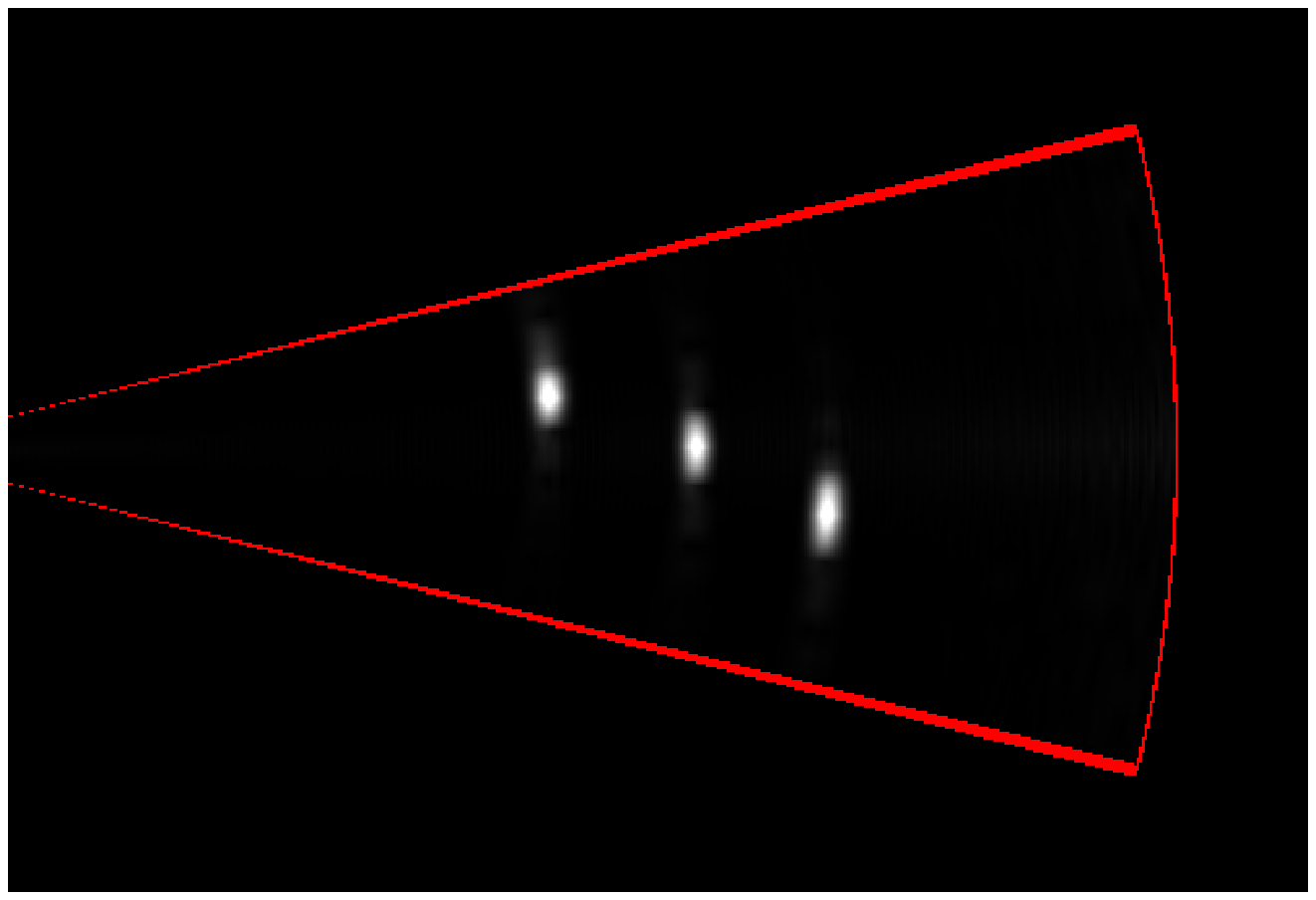}
        \vspace{-0.7cm}
        \subcaption{}
        \label{subfig:kWave Y=0 freq 100}
    \end{minipage}
    \hspace{-0.4cm}
    \hfill
    \begin{minipage}[b]{.2\linewidth}
        \centering \includegraphics[width = 5.5cm, angle = 270]{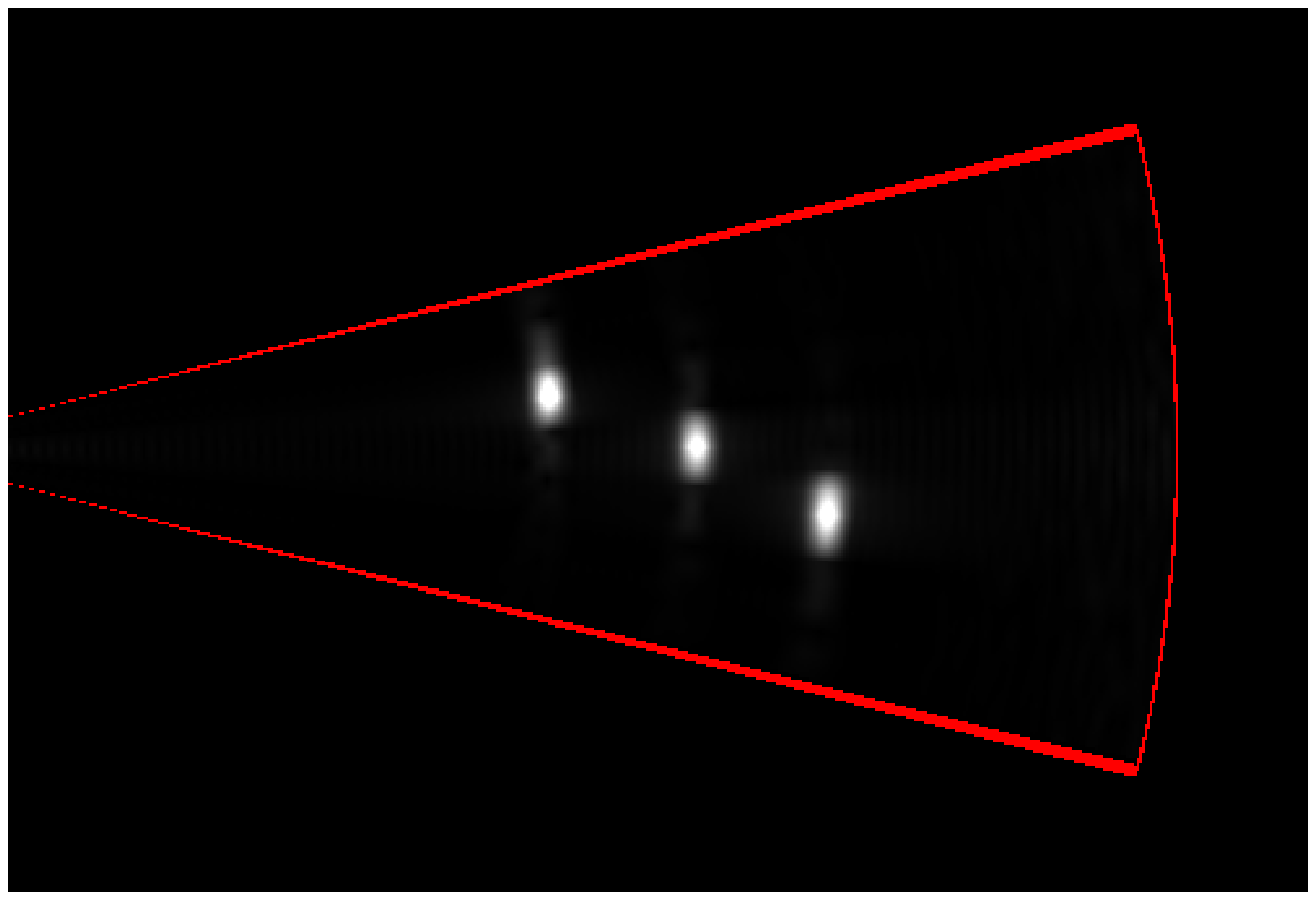}
        \vspace{-0.7cm}
        \subcaption{}
        \label{subfig:kWave Y=0 freq 50}
    \end{minipage}
    \hspace{-0.4cm}
    \hfill
    \begin{minipage}[b]{.2\linewidth}
        \centering \includegraphics[width = 5.5cm, angle = 270]{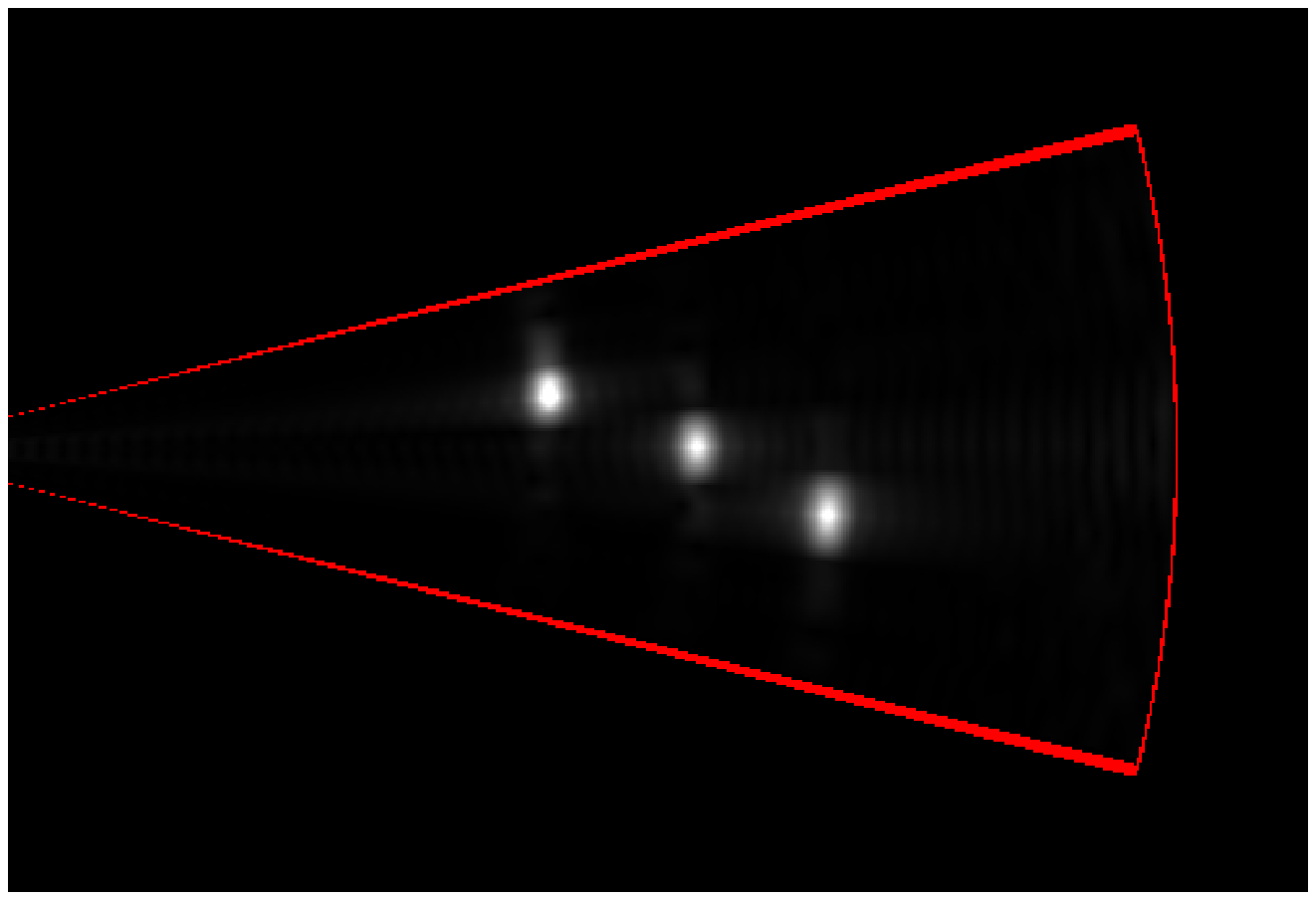}
        \vspace{-0.7cm}
        \subcaption{}
        \label{subfig:kWave Y=0 freq 33}
    \end{minipage}

    \vspace{0cm}

    \begin{minipage}[b]{.2\linewidth}
        \centering \includegraphics[width = 5.5cm, angle = 270]{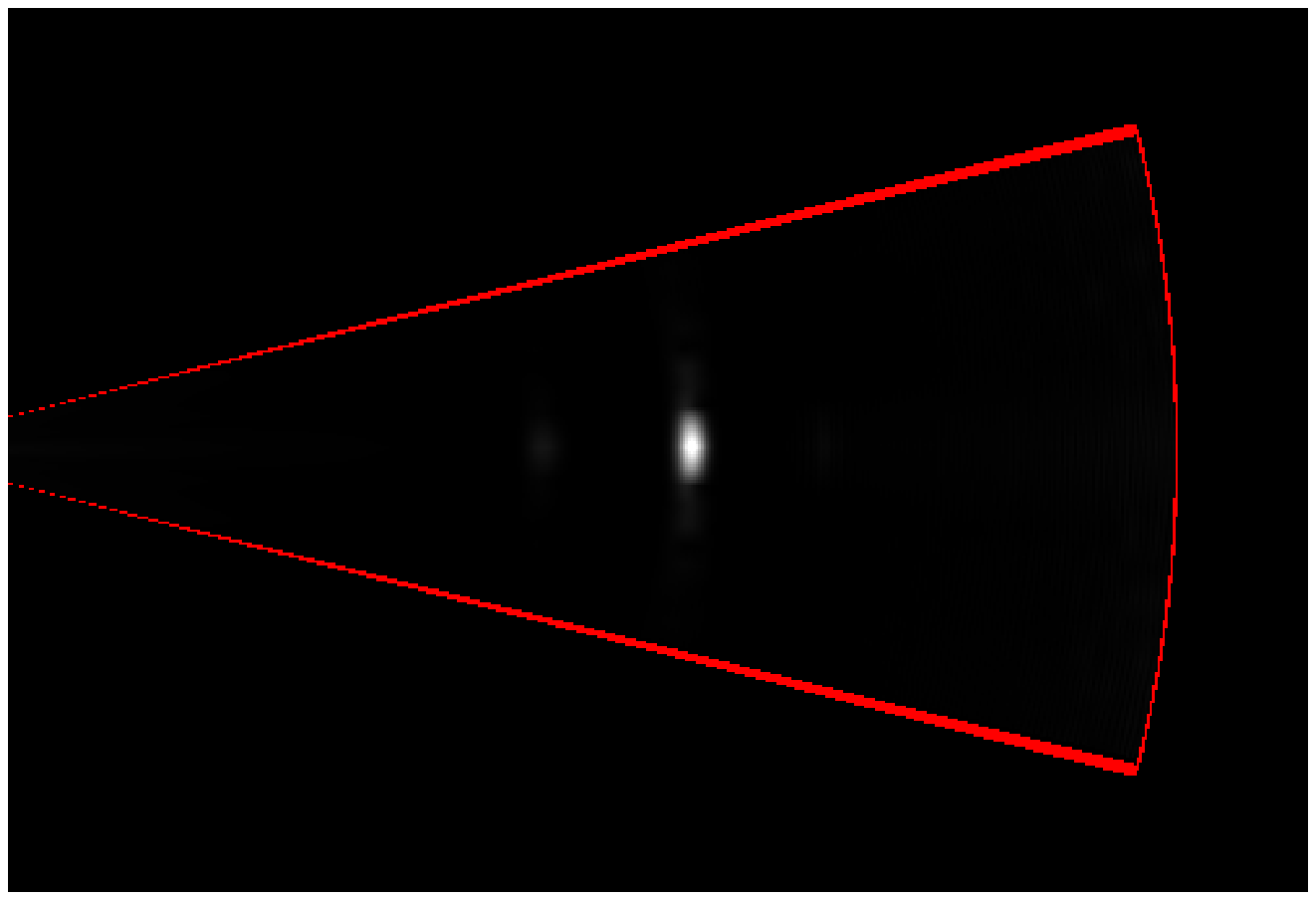}
        \vspace{-0.7cm}
        \subcaption{}
        \label{subfig:kWave X=0 time full}
    \end{minipage}
    \hspace{-0.4cm}
    \hfill
    \begin{minipage}[b]{.2\linewidth}
        \centering \includegraphics[width = 5.5cm, angle = 270]{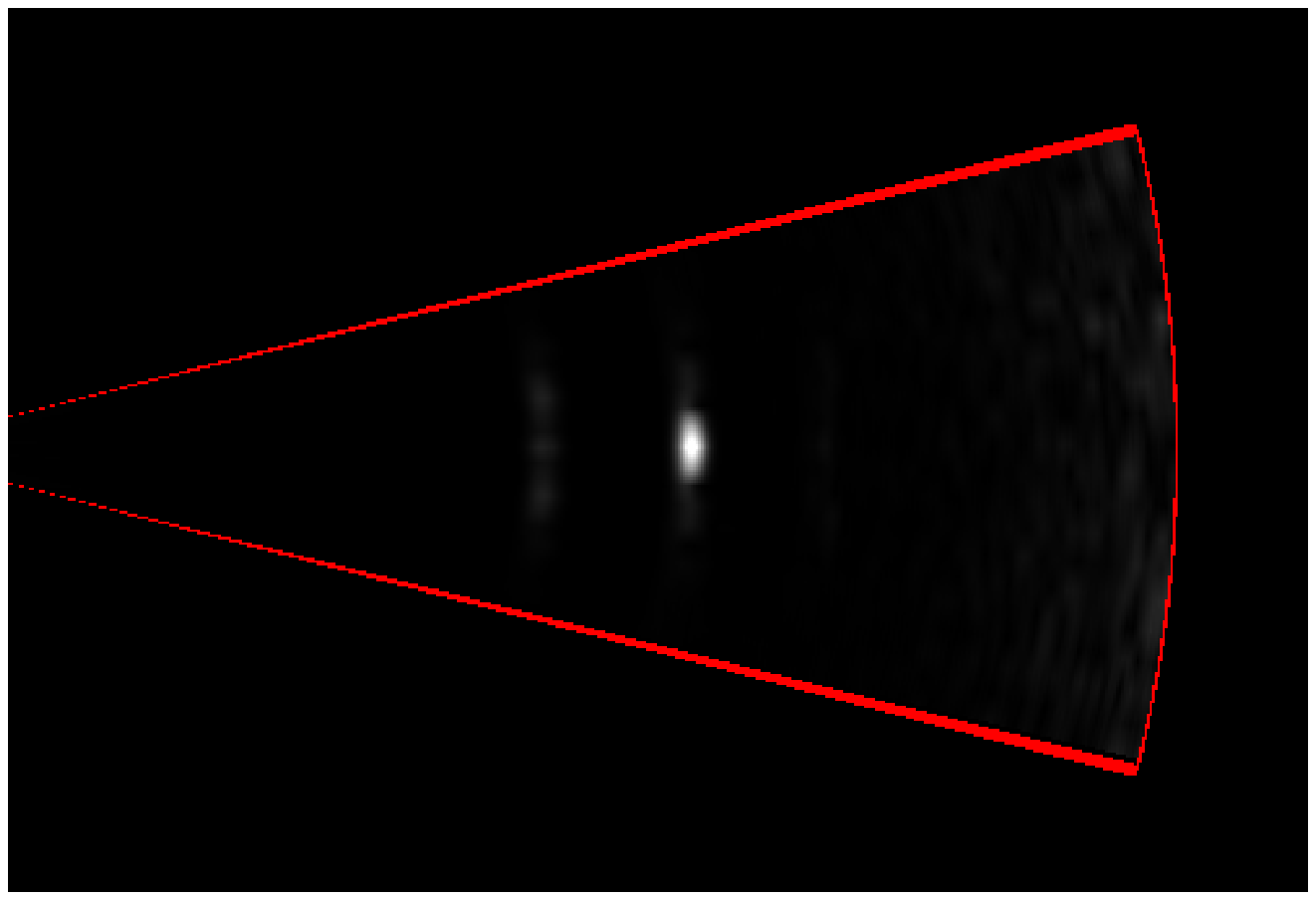}
        \vspace{-0.7cm}
        \subcaption{}
        \label{subfig:kWave X=0 time diagonals}
    \end{minipage}
    \hspace{-0.4cm}
    \hfill
    \begin{minipage}[b]{.2\linewidth}
        \centering \includegraphics[width = 5.5cm, angle = 270]{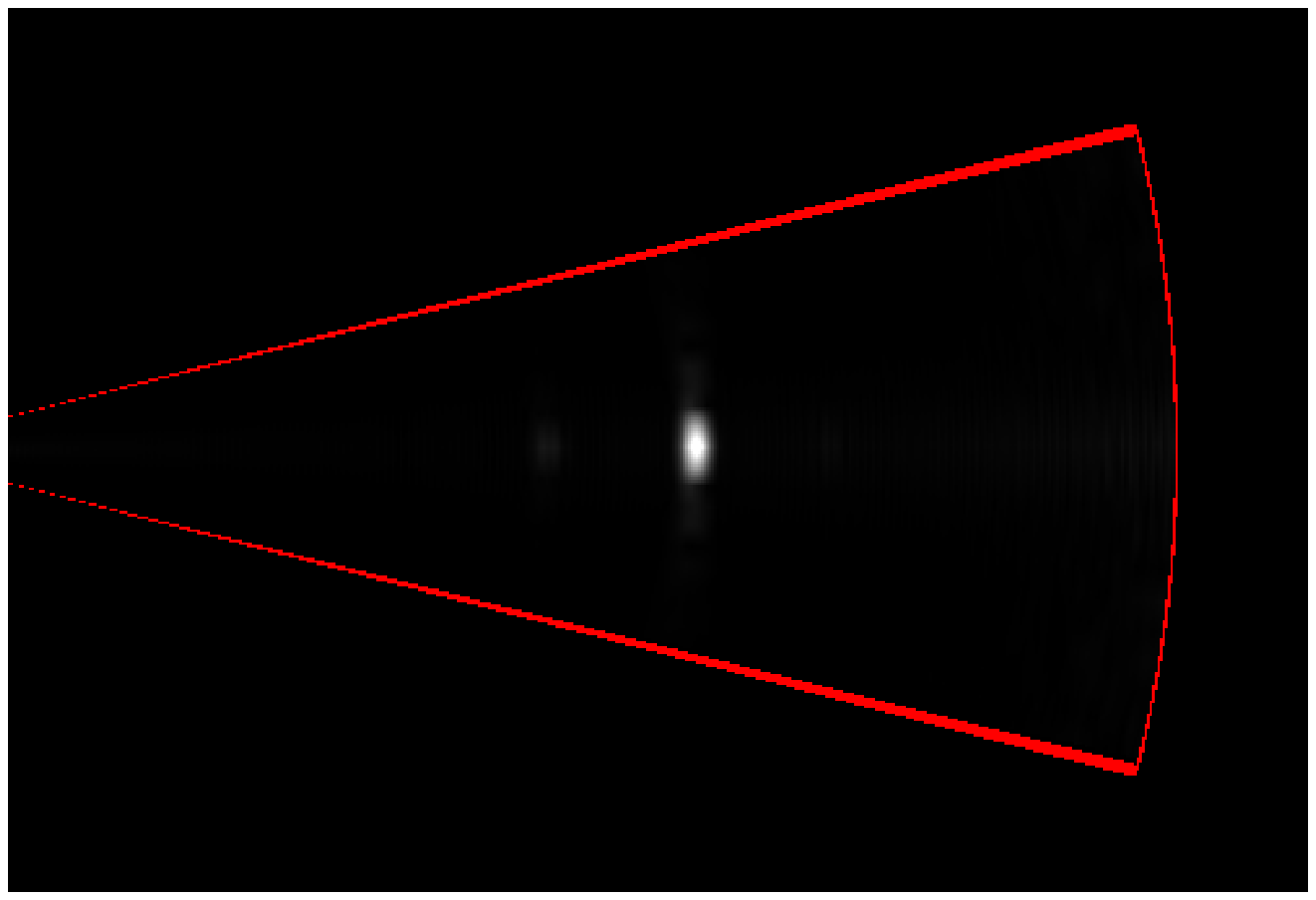}
        \vspace{-0.7cm}
        \subcaption{}
        \label{subfig:kWave X=0 freq 100}
    \end{minipage}
    \hspace{-0.4cm}
    \hfill
    \begin{minipage}[b]{.2\linewidth}
        \centering \includegraphics[width = 5.5cm, angle = 270]{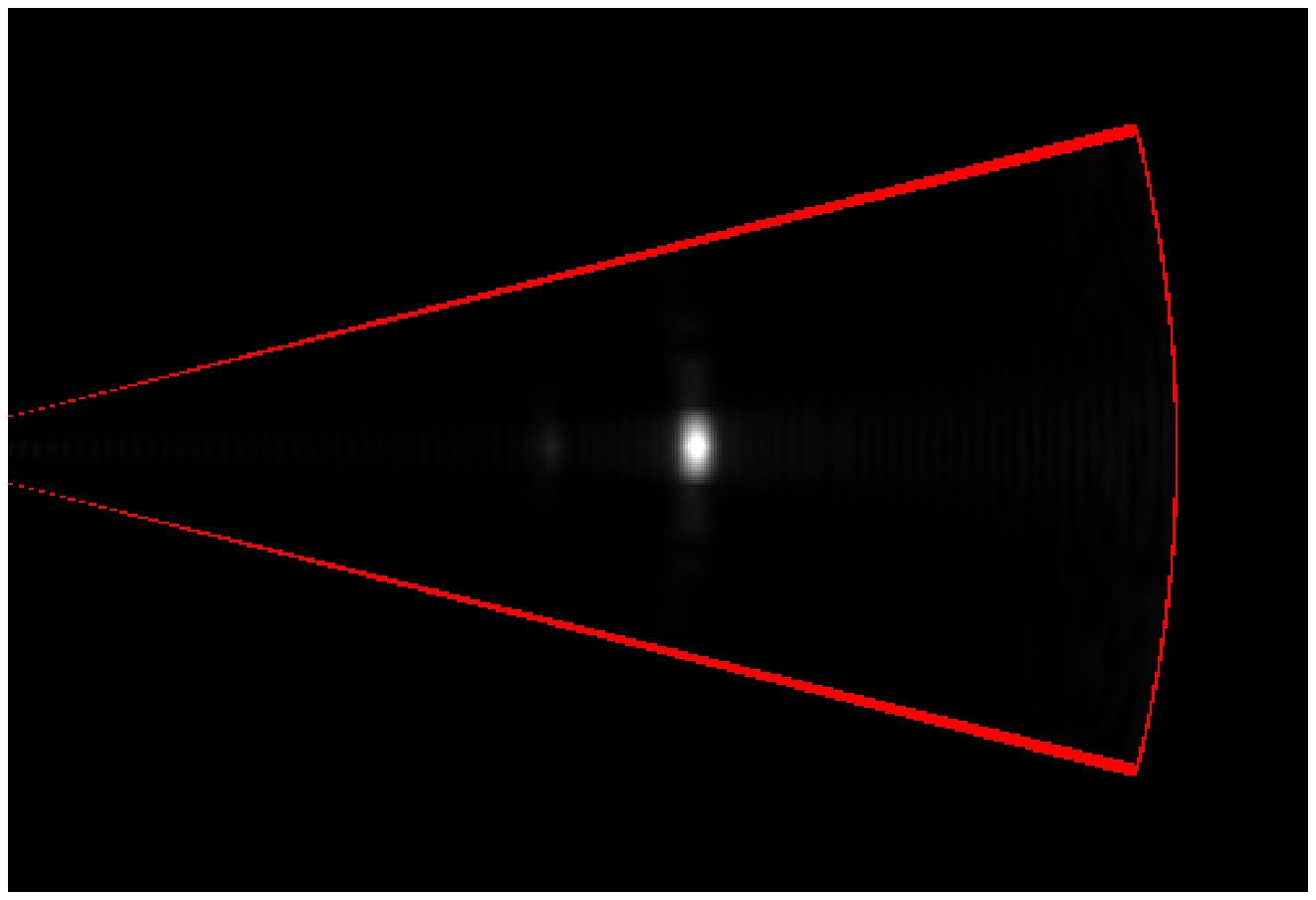}
        \vspace{-0.7cm}
        \subcaption{}
        \label{subfig:kWave X=0 freq 50}
    \end{minipage}
    \hspace{-0.4cm}
    \hfill
    \begin{minipage}[b]{.2\linewidth}
        \centering \includegraphics[width = 5.5cm, angle = 270]{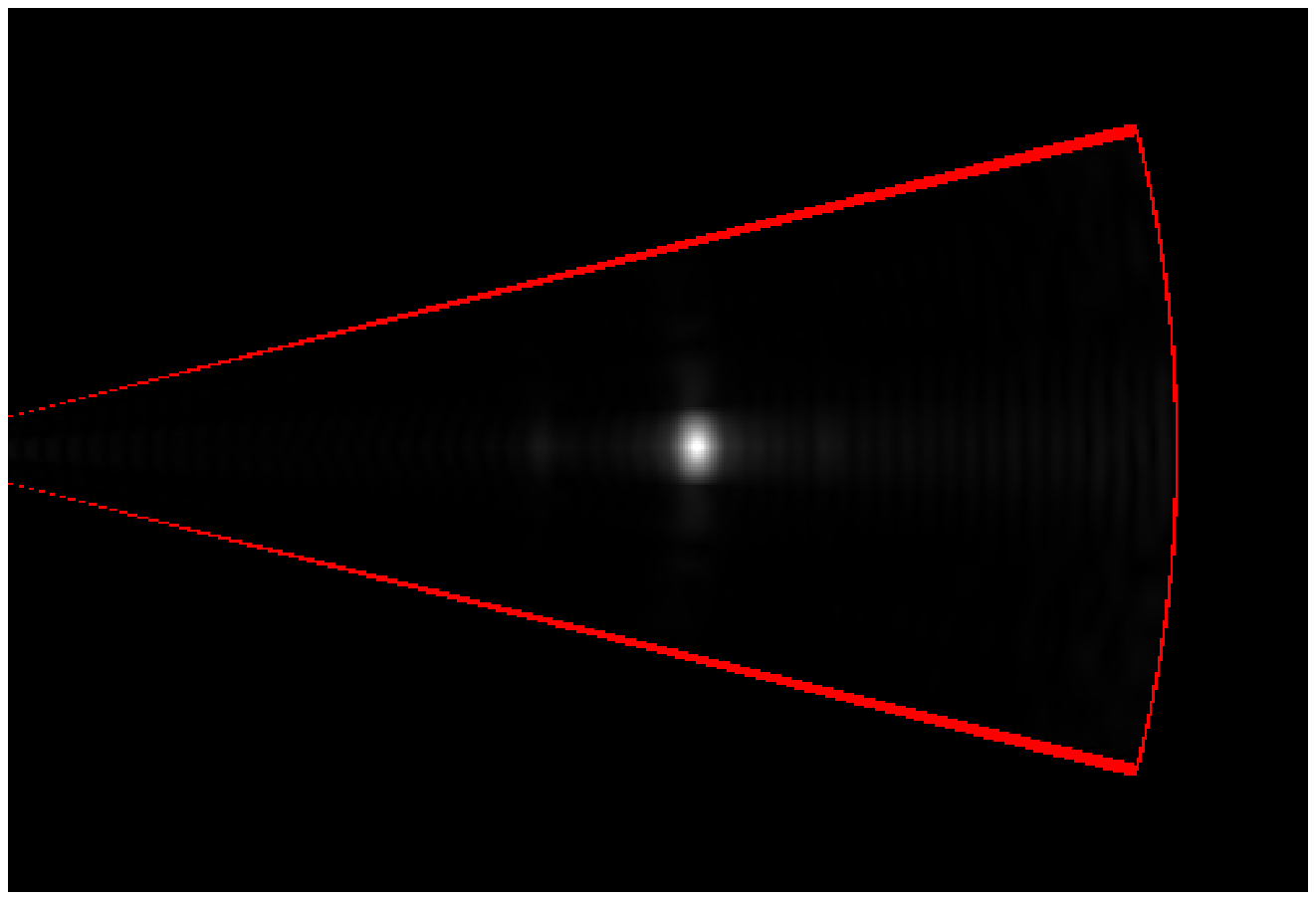}
        \vspace{-0.7cm}
        \subcaption{}
        \label{subfig:kWave X=0 freq 33}
    \end{minipage}
    \caption{Cross-sections of the simulated 3D imaging of three point reflectors placed on a plane. (a)-(e) display the $\theta_y=0^{\circ}$ cross-section, while (f)-(j) display the $\theta_x=0^{\circ}$ cross-section. (a), (f) display images acquired with time-domain beamforming, using the entire transducer grid. (b), (g)  display images acquired with time-domain beamforming, using the diagonals of the transducer grid. (c) and (h), (d) and (i), (e) and (j) display images acquired with FDBF using $B$, $B / 2$ and $B / 3$ DFT coefficients of the beamformed signal, respectively.}
    \label{fig:sim cross-sections}
\end{figure*}

Cross-sections of the resulting 3D volumes are displayed in Fig. \ref{fig:sim cross-sections}. It can be seen that all reflectors are clearly seen for all processing methods displayed. We note that the frequency-domain beamformed images display lower noise levels than the time-domain beamformed image reconstructed using a partial set of the transducer grid. The advantage of FDBF in the presence of noise is discussed in detail in Section \ref{subsec:sim noise}.

\begin{table}[htb]
  \begin{minipage}[b]{1.0\linewidth}
      \centering
      \vspace{0cm}
      \centerline{
        \begin{tabular}{|c|c|}
          \hline
          Processing method & Number of samples\\
          \hline \hline
          Time & \multirow{2}{*}{$5.89 \cdot 10^8$} \\
          Full grid & \\
          \hline
          Time & \multirow{2}{*}{$36.81\cdot10^6$} \\
          Diagonal grid & \\
          \hline
          Frequency & \multirow{2}{*}{$99.8\cdot10^6$} \\
          $B$ coefficients & \\
          \hline
          Frequency & \multirow{2}{*}{$54.64\cdot10^6$} \\
          $B / 2$ coefficients & \\
          \hline
          Frequency & \multirow{2}{*}{$39.74\cdot10^6$}\\
          $B / 3$ coefficients & \\
          \hline
        \end{tabular}}
      \vspace{-0.3cm}
  \end{minipage}
  \hfill
  \caption{Number of samples per volume for each processing method.}
  \label{tab:num of samples}
\end{table}

\subsection{Lateral and Axial Point Spread Functions}
\label{ssec:point spread functions}
\begin{figure}[htb]
    \vspace{-0.2cm} \hspace{0.2cm}
    \centering \includegraphics[width = \linewidth, angle = 0]{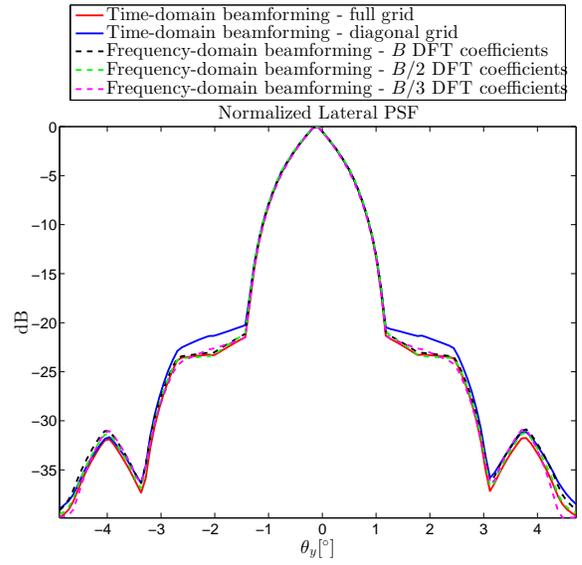}
    \vspace{-0.7cm}
    \caption{Normalized LPSFs for various processing methods.}
    \label{fig:LPSF}
\end{figure}

\begin{figure}
    \centering \includegraphics[width = \linewidth, angle = 0]{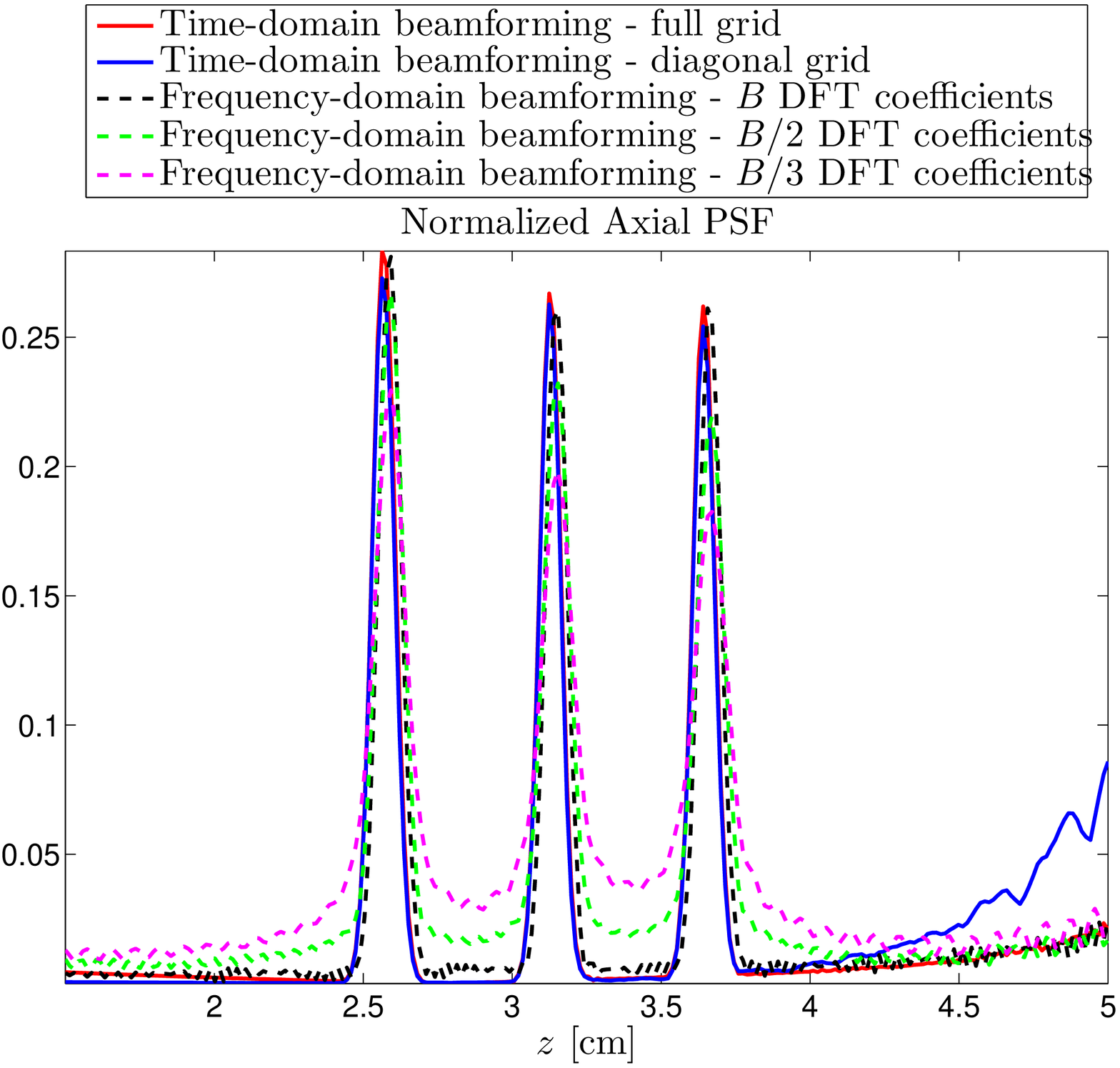}
    \caption{Normalized APSFs for various processing methods.}
    \label{fig:APSF}
    \vspace{-0.2cm}
\end{figure}

We next compare our proposed method to time-domain beamforming by calculating the LPSF and APSF characterizing each processing method.
The LPSFs are acquired for the reflector placed at the transmit focus point and plotted on constant-$r$ arcs. 
The APSFs show the sum of the constant-$\theta_x$ and $\theta_y$ lines on which point reflectors are placed. The LPSFs are normalized such that the maximum at $\theta_y=0^{\circ}$ is set to 1, while the APSFs are normalized to unit energy. The PSFs are presented in Figs.~\ref{fig:LPSF} and \ref{fig:APSF}.

\begin{table}[htb]
  \begin{minipage}[b]{1.0\linewidth}
      \centering
      \centerline{
        \begin{tabular}{|c|c|c|c|}
          \hline
          Processing & Full width at & Average & First side- \\
          method & half maximum & side-lobes & lobe's peak \\
          \hline \hline
          Time& \multirow{2}{*}{$1.50^{\circ}$} & \multirow{2}{*}{$-29.63$ dB} & \multirow{2}{*}{$-22.93$ dB} \\
          Full grid & & & \\
          \hline
          Time& \multirow{2}{*}{$1.51^{\circ}$} & \multirow{2}{*}{$-28.53$ dB} & \multirow{2}{*}{$-21.40$ dB}
          \\
          Diagonal grid& & & \\
          \hline
          Frequency & \multirow{2}{*}{$1.51^{\circ}$} & \multirow{2}{*}{$-29.01$ dB} & \multirow{2}{*}{$-22.70$ dB}
          \\
          $B$ coefficients & & & \\
          \hline
          Frequency & \multirow{2}{*}{$1.52^{\circ}$} & \multirow{2}{*}{$-29.37$ dB} & \multirow{2}{*}{$-22.90$ dB}
          \\
          $B / 2$ coefficients & & & \\
          \hline
          Frequency & \multirow{2}{*}{$1.47^{\circ}$} & \multirow{2}{*}{$-29.47$ dB} & \multirow{2}{*}{$-22.74$ dB}
          \\
          $B / 3$ coefficients & & &
          \\
          \hline
        \end{tabular}}
      \vspace{-0.3cm}
  \end{minipage}
  \hfill
  \caption{LPSFs properties.}
  \label{tab:LPSF table}
\end{table}

\begin{table}[htb]
  \begin{minipage}[b]{1.0\linewidth}
      \centering
      \centerline{
        \begin{tabular}{|c|c|}
          \hline
          Processing method & Full width at half maximum\\
          \hline \hline
          Time & \multirow{2}{*}{$0.084$ mm} \\
          Full grid & \\
          \hline
          Time & \multirow{2}{*}{$0.082$ mm} \\
          Diagonal grid & \\
          \hline
          Frequency & \multirow{2}{*}{$0.086$ mm} \\
          $B$ coefficients & \\
          \hline
          Frequency & \multirow{2}{*}{$0.107$ mm} \\
          $B / 2$ coefficients & \\
          \hline
          Frequency & \multirow{2}{*}{$0.138$ mm}\\
          $B / 3$ coefficients & \\
          \hline
        \end{tabular}}
      \vspace{-0.3cm}
  \end{minipage}
  \hfill
  \caption{APSFs properties.}
  \label{tab:APSF table}
\end{table}

The properties of the LPSFs are displayed in Table \ref{tab:LPSF table}. We see that the LPSFs obtained with 3D FDBF for different rate reduction factors display very similar properties to the LPSF acquired
with time-domain beamforming using the full transducer grid, and exhibit improved results over the LPSF acquired with time-domain beamforming using only the diagonals of the grid. This is an expected result, since our method reconstructs the axial lines of the image and does not have a direct effect on the lateral resolution.
The widths of the peak located at the focus point, acquired for each reconstructed method, are shown in Table \ref{tab:APSF table}. As seen in Table \ref{tab:APSF table} and Fig. \ref{fig:APSF}, the APSFs deteriorate when the amount of Fourier coefficients used to reconstruct the image is decreased. We note that energy leakage from the peaks is increased when less Fourier coefficients of the beamformed signal are used in the reconstruction process. However the effect of APSF deterioration becomes visible only when less then half the bandwidth is used for signal reconstruction.

Reducing the amount of transducer elements enhances noise levels in the image and deteriorates the lateral resolution, while our proposed reconstruction method affects mainly the axial resolution. Acknowledging this fact, we may consider a midway approach, where rate reduction is achieved both by reducing the amount of transducer elements and applying frequency-domain beamforming. The more dominant factor for rate reduction will be dictated according to the trade-off stated above.
In addition to the axial and lateral resolution, another important feature that has to be regarded is the SNR. It can be seen in Fig. \ref{fig:APSF} that the line acquired with time-domain beamforming using partial grid data contains high levels of noise at the far-field, since less transducer elements participate in the delay-and-sum process. This holds even when no noise was incorporated in the simulation - the ``effective noise" stems from reflections corresponding to the numerical solution of the simulation code.

\subsection{Simulation with Noise}
\label{subsec:sim noise}
To show the advantage of our proposed method over the partial grid time-domain beamforming method in terms of SNR, another simulation is conducted. A pulse is transmitted in the $\theta_x=0,~\theta_y=0$ direction. A single large reflector is placed at the focus depth of the transmitted pulse. The signals detected at the transducer elements are contaminated by white Gaussian noise imitating the thermal noise of the system. We then proceed to reconstruct the $\theta_x=0,~\theta_y=0$ beam using all five methods described in Section \ref{ssec:simulation setup}. In addition, clean beams, without the addition of noise, are obtained for all five methods. We denote the noisy and clean beamformed lines by $\Phi_{\textrm{noise}}(t)$ and $\Phi_{\textrm{clean}}(t)$, respectively.

We define the SNR of a beam as the ratio between the energy stored in a segment of length $5\lambda$ around the main peak of the beam, where $\lambda$ is the wavelength corresponding to the carrier frequency of the transmitted pulse, and the energy of the noise in the beamformed line, defined as $n(t) = \Phi_{\textrm{noise}}(t)-\Phi_{\textrm{clean}}(t)$. That is:
\begin{equation}
\label{SNR}
\vspace{-0.1cm}
\mathrm{SNR} = 10\log_{10} \left( \frac { \int |\Phi_{\textrm{clean}}(t)|^2 dt } { \int |n(t)|^2 dt } \right).
\end{equation}
The results are displayed in Table \ref{tab:SNR table}. As expected, the reduction in number of elements participating leads to a dramatic reduction in SNR. In contrast, 3D FDBF displays higher SNR even over the time-domain processing when the entire grid of transducer elements is used. A remarkable point is that the SNR increases when less Fourier coefficients are involved in the reconstruction of the beam. This is not surprising since the noise is equally spaced over the entire spectrum of the system -- the fewer Fourier coefficients used in the reconstruction process, the less noise involved.

To conclude, 3D FDBF using half the bandwidth is comparable to standard time domain processing with diagonal transducer elements in terms of data rate reduction, but outperforms it by 20 dB in terms of SNR.
\begin{table}[htb]
  \begin{minipage}[b]{1.0\linewidth}
      \centering
      \centerline{
        \begin{tabular}{|c|c|}
          \hline
          Processing method & SNR\\
          \hline \hline
          Time & \multirow{2}{*}{20.83 dB} \\
          Full grid & \\
          \hline
          Time & \multirow{2}{*}{8.92 dB} \\
          Diagonal grid & \\
          \hline
          Frequency & \multirow{2}{*}{26.69 dB} \\
          $B$ coefficients & \\
          \hline
          Frequency & \multirow{2}{*}{28.68 dB} \\
          $B / 2$ coefficients & \\
          \hline
          Frequency & \multirow{2}{*}{29.48 dB}\\
          $B / 3$ coefficients & \\
          \hline
        \end{tabular}}
  \end{minipage}
  \hfill
  \caption{SNR of processing methods.}
  \label{tab:SNR table}
\end{table}
%


\subsection{Application on a Commercial System}
\label{subsec:commercial system}

\begin{figure*}
    \centering
    \begin{minipage}[b]{.45\linewidth}
        \centering
        \includegraphics[width = 0.75\linewidth]{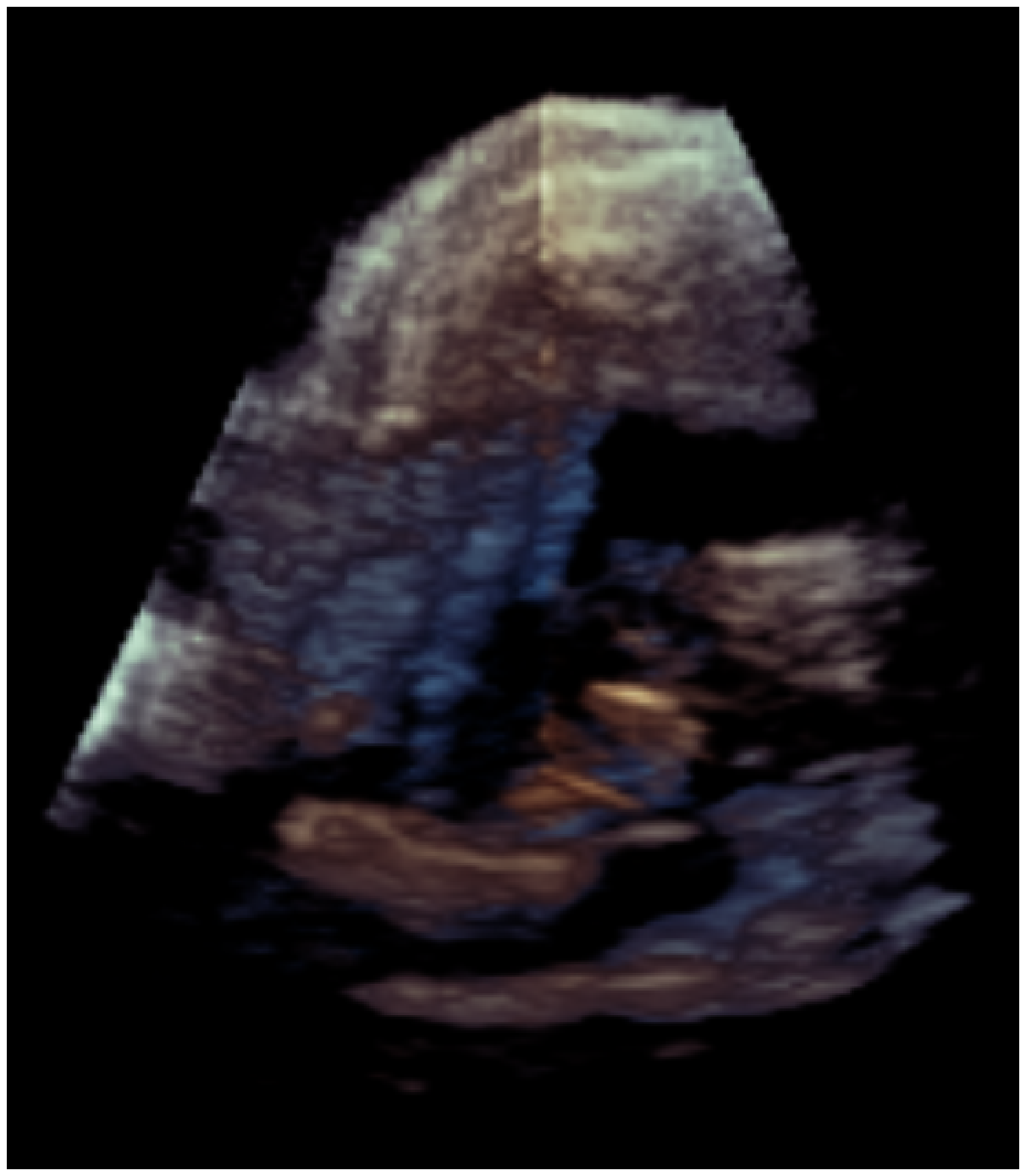}
        \subcaption{}
        \label{subfig:comm time frame}
    \end{minipage}
    \begin{minipage}[b]{.45\linewidth}
        \centering
        \includegraphics[width = 0.75\linewidth]{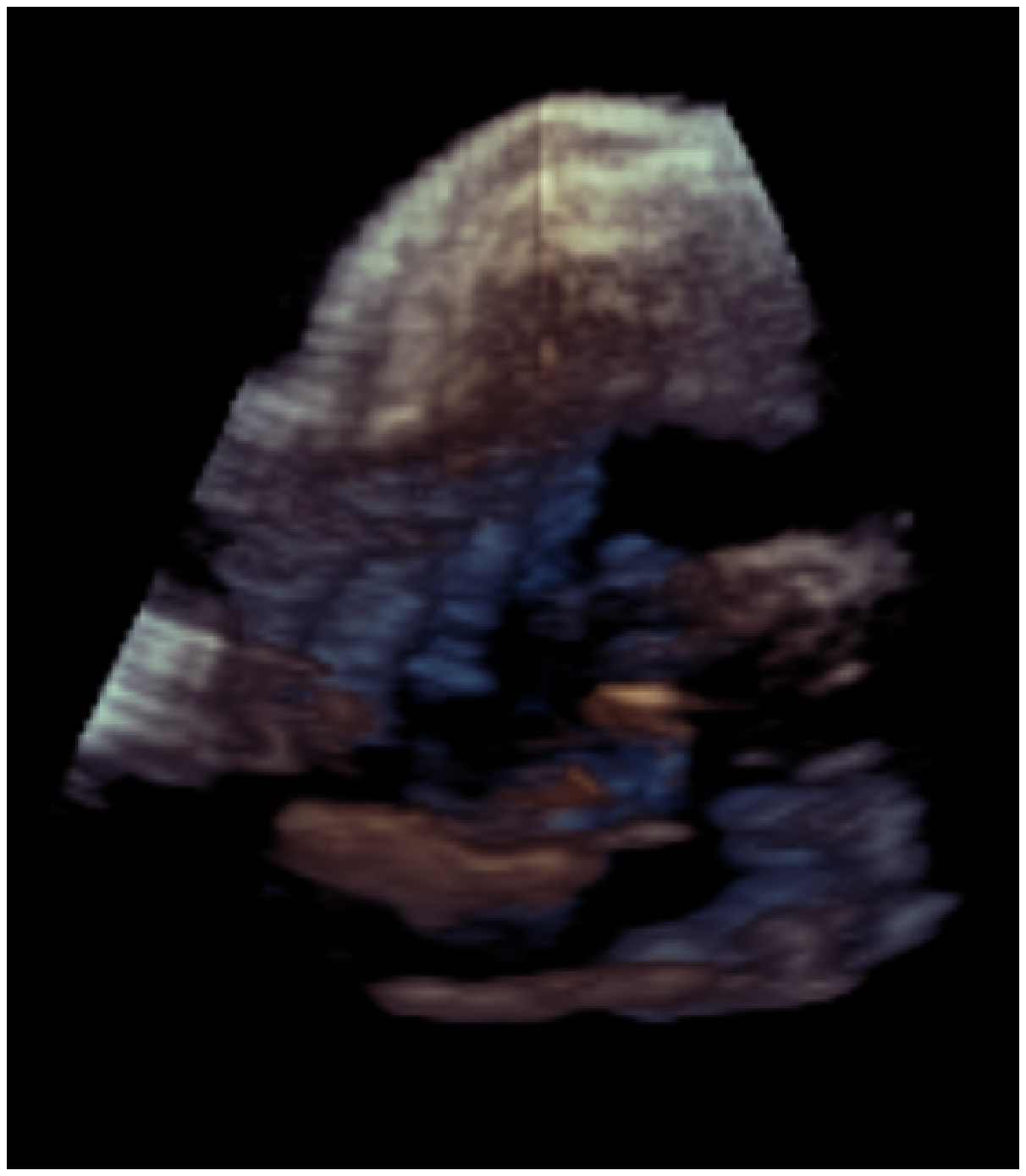}
        \subcaption{}
        \label{subfig:comm half freq frame}
    \end{minipage}
    \caption{3D imaging of a phantom of a heart ventricle. (a) displays the time-domain reconstruction of the frame, while (b) displays the frequency-domain reconstructed frame, using $\mathcal{K}=B/2$ Fourier coefficients of the beamformed signal with 12 fold rate reduction.}
    \label{fig:comm volumetric}
\end{figure*}

We now demonstrate our method on data collected using a commercial 3D ultrasound system, while imaging a phantom of a heart ventricle. Shown in Fig. \ref{fig:comm volumetric} are images of the entire volumetric scan, taken from a specific angle, for demonstration. The frames are reconstructed using time-domain beamforming and frequency-domain beamforming with $\mathcal{K}=B/2$ coefficients of the beamformed signal. The complex structure of the phantom allows us to test the performance of the proposed method on volumes containing multiple strong reflectors as well as weak reflectors known as speckle.

\begin{figure*}[htb]
    \begin{minipage}[b]{.33\linewidth}
        \centering \includegraphics[width = 5.5cm]{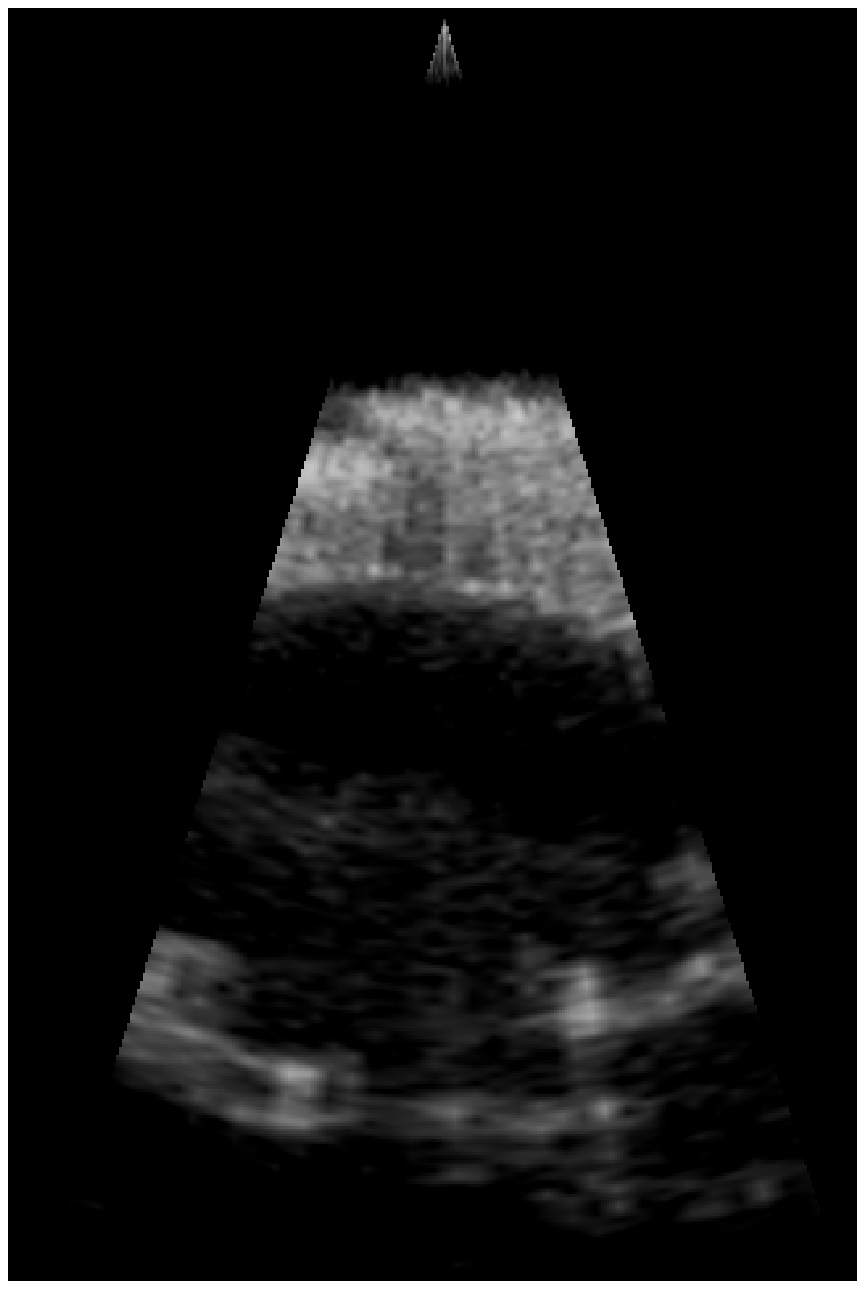}
        \subcaption{}
        \label{subfig:comm time 1}
    \end{minipage}
    \hspace{-0.4cm}
    \hfill
    \begin{minipage}[b]{.33\linewidth}
        \centering \includegraphics[width = 5.5cm]{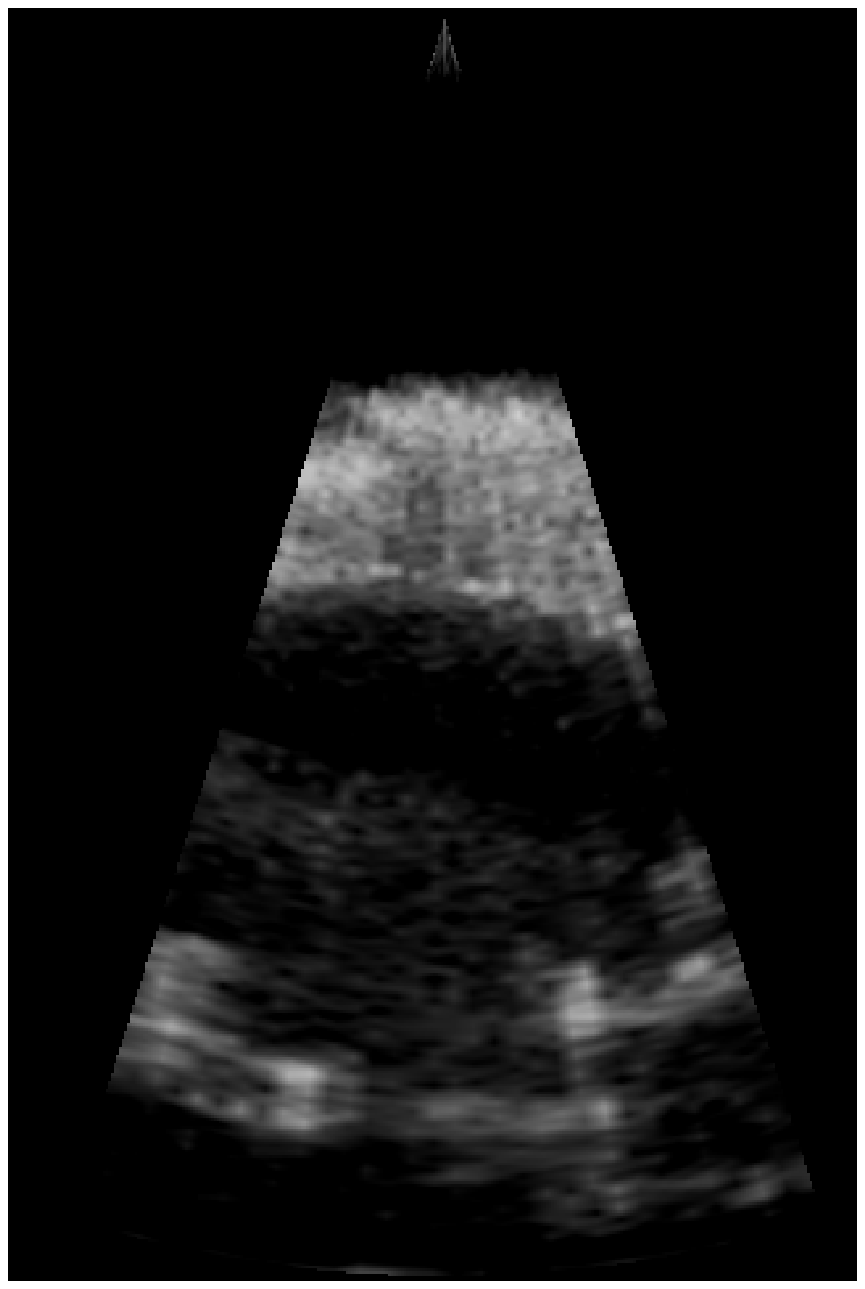}
        \subcaption{}
        \label{subfig:comm full freq 1}
    \end{minipage}
    \hspace{-0.4cm}
    \hfill
    \begin{minipage}[b]{.33\linewidth}
        \centering \includegraphics[width = 5.5cm]{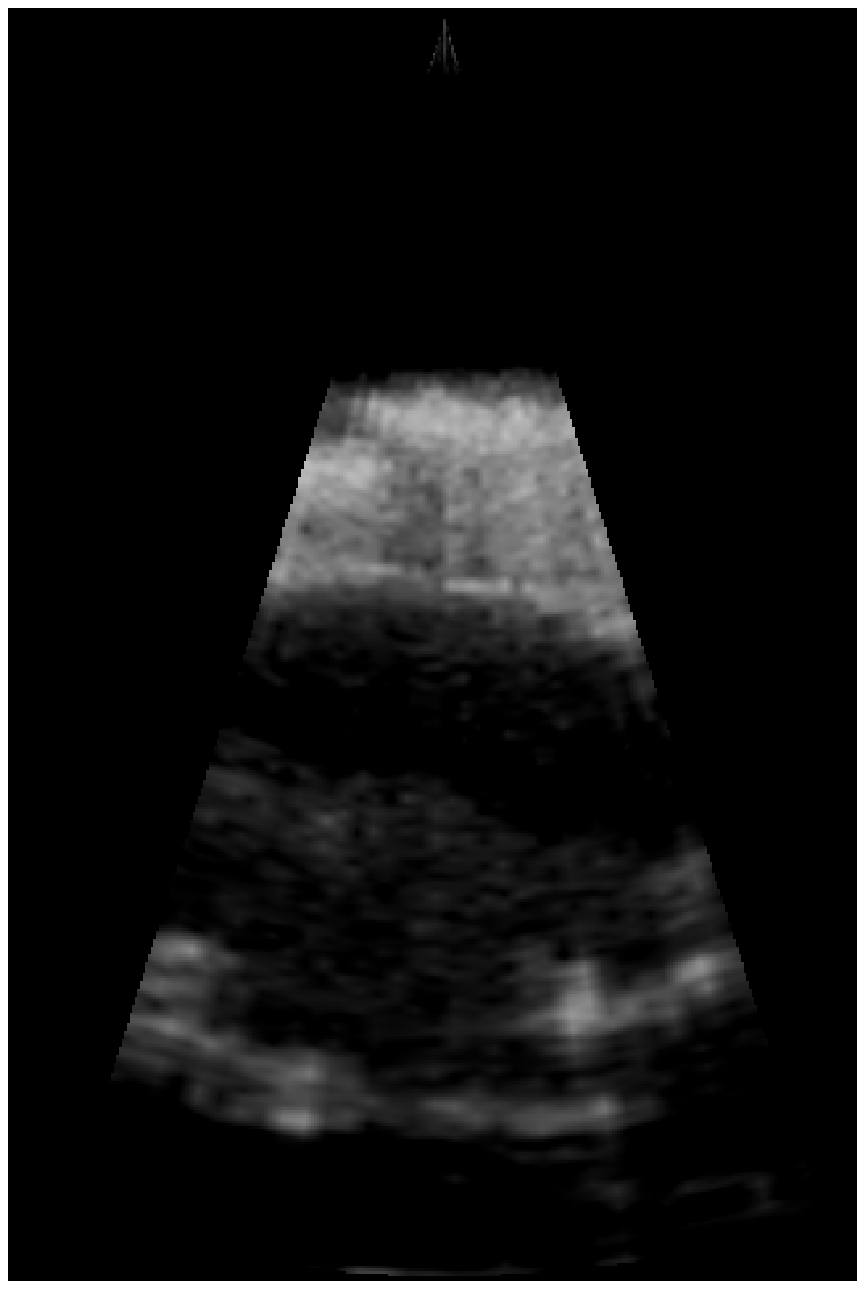}
        \subcaption{}
        \label{subfig:comm half freq 1}
    \end{minipage}

    \begin{minipage}[b]{.33\linewidth}
        \centering \includegraphics[width = 5.5cm]{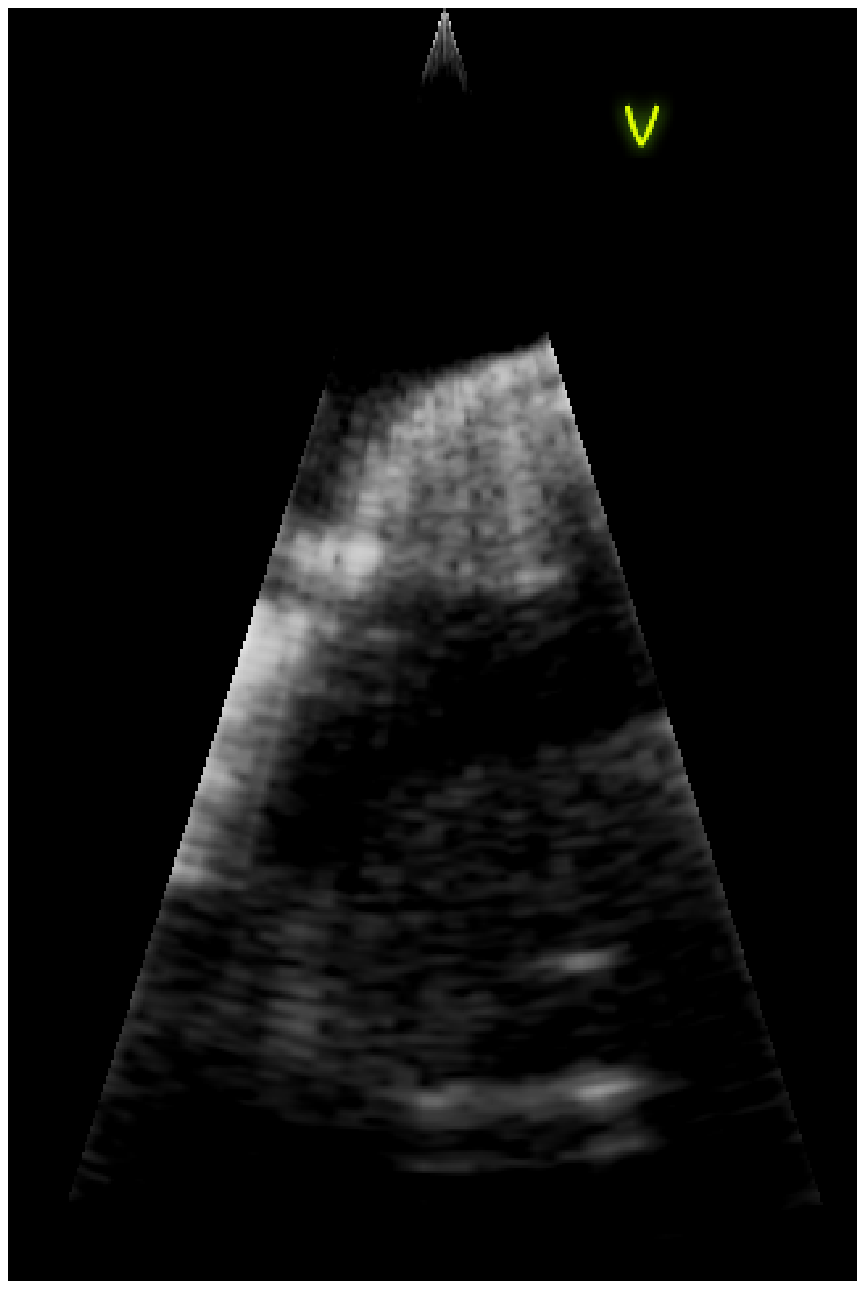}
        \subcaption{}
        \label{subfig:comm time 2}
    \end{minipage}
    \hspace{-0.4cm}
    \hfill
    \begin{minipage}[b]{.33\linewidth}
        \centering \includegraphics[width = 5.5cm]{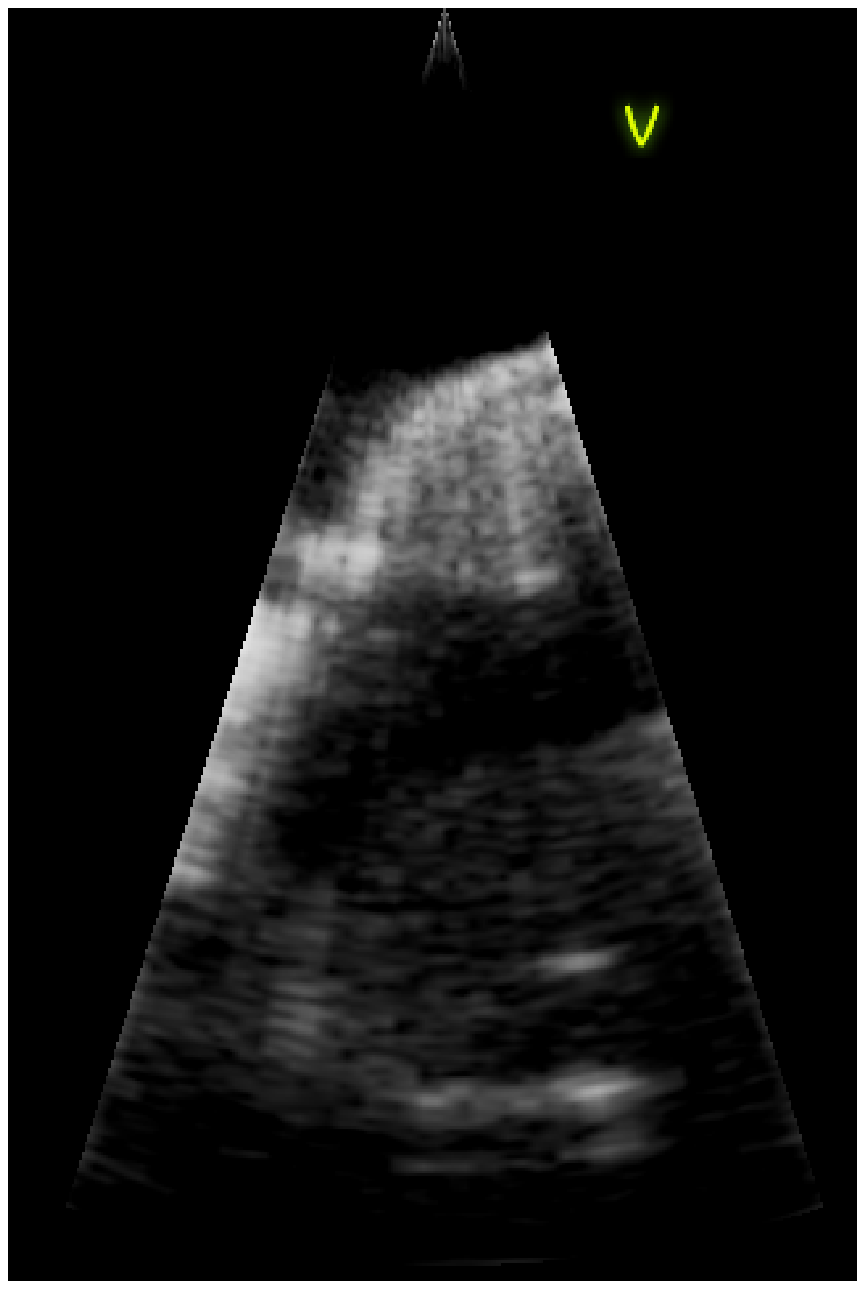}
        \subcaption{}
        \label{subfig:comm full freq 2}
    \end{minipage}
    \hspace{-0.4cm}
    \hfill
    \begin{minipage}[b]{.33\linewidth}
        \centering \includegraphics[width = 5.5cm]{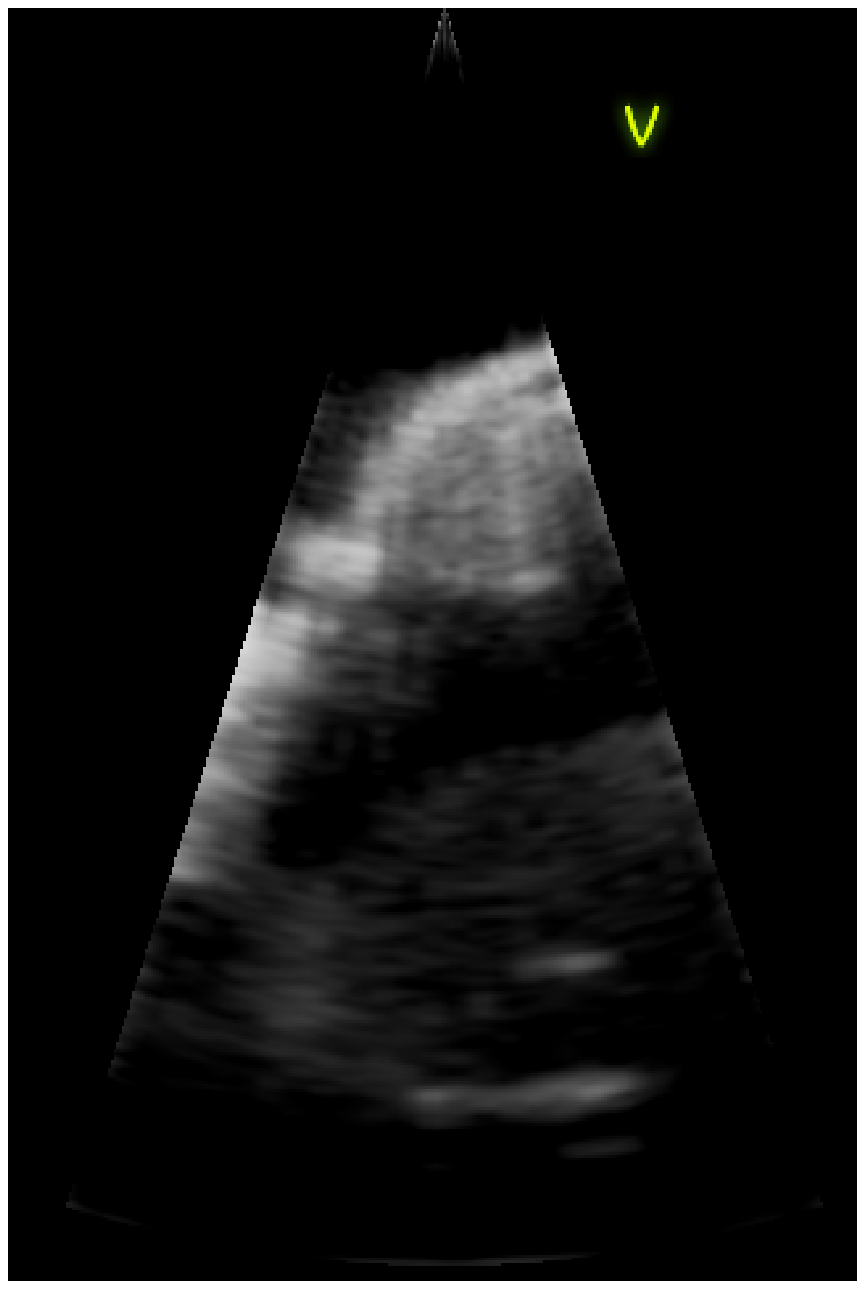}
        \subcaption{}
        \label{subfig:comm half freq 2}
    \end{minipage}
    \caption{Cross-sections of the 3D imaging of a phantom of a heart ventricle. (a)-(c) display the $\theta_y=0^{\circ}$ cross-section, while (d)-(f) display the $\theta_x=0^{\circ}$ cross-section. (a),(d) display images acquired using time-domain beamforming. (b), (e) and (c), (f) display images acquired with frequency-domain beamforming 
    with 6 and 12 fold rate reduction respectively.}
    \label{fig:Comm cross-sections}
\end{figure*}

The transducer grid is comprised of 2000 transducers. The entire grid participates in the transmission stage, while analog sub-array beamforming is performed in the reception stage. This sub-optimal processing method is required 
to adjust the number of elements to the number of electronic channels.
We processed the collected data in the same manner as in the previous section, using time-domain beamforming requiring 3120 samples per image line, and frequency domain beamforming for $\mathcal{K} = B $ and $\mathcal{K} = B/2$ with $B=506$. When $B$ Fourier coefficients are computed, the processing is performed at the signal's effective Nyquist rate and oversampling is avoided leading to 6 fold rate reduction. FDBF with $B/2$ coefficients implies 12 fold rate reduction leading to sampling and processing at a sub-Nyquist rate. 
Cross-sections of the 3D frames acquired are displayed in Fig. \ref{fig:Comm cross-sections}.

The image obtained by low-rate FDBF  for $\mathcal{K} = B $ is virtually identical to one obtained by standard time-domain processing at a high rate. This result is expected since for $\mathcal{K} = B $ all the information is obtained in frequency. We also note prominent similarity between the image obtained at the sub-Nyquist rate and the original one. In particular, the strong reflectors and the speckle pattern are preserved. 
The above results prove that FDBF can be combined with analog sub-array beamforming without significant reduction in image quality. In this way channel number reduction resulting from analog sub-array processing is combined with sampling rate reduction obtained by FDBF paving the way to real-time low-cost 3D imaging system.

%
\section{Conclusions}
\label{sec:conclusion}

In this work a solution to one of the major bottlenecks in 3D imaging, the amount of sampled data, is introduced.
The number of samples taken at each transducer element is reduced by applying the low-rate sampling scheme presented in \cite{chernyakova2013compressed} to the individual signals detected by the 2D grid. To benefit from the achieved data rate reduction we prove that the subsequent processing, namely, 3D beamforming, can be performed directly in frequency. 
The translation of beamforming to the frequency domain allows bypassing oversampling and to obtain $4-10$ fold rate reduction without any assumptions on the signal's structure.

When signal's structure is exploited further rate reduction is obtained. We prove that 3D beamformed signal obeys an FRI model, which allows us to sample and process the signals at sub-Nyquist rate while retaining sufficient image quality.

The performance of the proposed method is verified in terms of both LPSF and APSF. It is shown that in accordance with our expectation it has no effect on LPSF, while the APSF is virtually the same when the entire set of Fourier coefficients of the beam within its effective bandwidth is computed. For sub-Nyquist processing APSF is slightly reduced, however when half the beam's bandwidth is used, the degradation is negligible. We also demonstrate the advantage of 3D FDBF in the presence of noise. The simulations with noise show that low-rate 3D FDBF outperforms not only the time-domain processing with the partial grid of elements participating in reception, but also the optimal time-domain processing with a full grid.

Finally we incorporate the proposed framework to a commercial imaging system and combine it with analog sub-array beamforming, required to adjust the number of elements to the number of electronic channels. The results verify that no further image degradation is introduced and that our approach can be used in conjunction with spatial sub-sampling techniques.

Our results pave the way for low-cost real-time capability crucial for further development of 3D ultrasound imaging.

\begin{appendices}
\section{FRI Model of the Beamformed Signal}
\label{sec:app beamformed FRI}
We assumed in equation \eqref{individual FRI} that the individual signals obey the FRI model. We wish to prove that the beamformed signal approximately obeys the FRI model, so that \eqref{beam FRI} holds.

In order to show this, we rely on three reasonable assumptions.
First, we assume that $2(|\gamma_m|+|\gamma_n|+|\gamma_{m,n}^z|) \le t_l$. Such a constraint may be forced by applying time-dependent apodization, in such a way that $\varphi(t;\theta_x,\theta_y)$ is omitted from the delay and sum process in \eqref{phi beamformed} for $t \ge 2(|\gamma_m| + |\gamma_n| + |\gamma_{m,n}^z|)$.
Second, we assume that the pulse $h(t)$, transmitted to the medium from each of the individual transducer elements and reflected back from scatterers in the medium, is compactly supported on the interval $[0,\Delta)$.
Finally, we assume $\Delta \ll t_l$. Again, such a constraint may be forced by applying apodization.

Using \eqref{individual FRI}, the individual distorted signals in \eqref{phi beamformed} are of the form
\begin{equation}\label{individual distorted}
\vspace{-0.1cm}
\hat{\varphi}_{m,n}(t;\theta_x,\theta_y) = \sum_{l=1}^L \tilde{a}_{l,m,n} h(\tau_{m,n}(t;\theta_x,\theta_y) - t_{l,m,n}).
\vspace{-0.1cm}
\end{equation}
The resulting signal comprises $L$ pulses, which are distorted versions of the pulse $h(t)$.

Suppose that some of these pulses originated in reflectors located off the central beam axis. When we combine the individual signals in \eqref{phi beamformed} to calculate the beamformed signal, these pules will be attenuated due to destructive interference. Therefore, when considering the beamformed signal $\Phi(t;\theta_x,\theta_y)$, we are concerned only with the pulses which originated in reflectors located along the central beam axis. For convenience, we assume that all pulses in \eqref{individual distorted} satisfy this property - those that do not will vanish in $\Phi(t;\theta_x,\theta_y)$.

We note that the time of arrival at the $(m,n)$ element, $t_{l,m,n}$, is related to the time of arrival at the $(m_0,n_0)$ element according to the alignment introduced in \eqref{phim}. Thus, we can express the delays of the individual signals, $\{t_{l,m,n}\}_{l=1}^L$, in terms of $t_l$, as
\begin{equation}\label{distorted delays}
\vspace{-0.1cm}
t_{l,m,n} = \tau_{m,n}(t_l;\theta_x,\theta_y).
\vspace{-0.1cm}
\end{equation}
Therefore, we may rewrite \eqref{individual distorted} as
\begin{equation}\label{distorted delays 2}
\vspace{-0.1cm}
\hat{\varphi}_{m,n}(t;\theta_x,\theta_y) = \sum_{l=1}^L {\tilde{a}_{l,m,n} \tilde{h}(t;\theta_x,\theta_y)},\vspace{-0.1cm}
\end{equation}
where we defined $\tilde{h}(t;\theta_x,\theta_y)=h(\tau_{m,n}(t;\theta_x,\theta_y) - \tau_{m,n}(t_l;\theta_x,\theta_y))$.

Applying our second assumption, we find that the support of $\tilde{h}(t;\theta_x,\theta_y)$ is defined by the requirement
\begin{equation}\label{distorted pulse support 1}
\vspace{-0.1cm}
0 \le \tau_{m,n}(t;\theta_x,\theta_y) - \tau_{m,n}(t_l;\theta_x,\theta_y) < \Delta.\vspace{-0.1cm}
\end{equation}
It can be shown that the inequalities in \eqref{distorted pulse support 1} are satisfied for $t \in [t_l, t_l + \Delta')$, where
\begin{equation}\label{distorted pulse support 2}
\vspace{0cm}
\Delta' = 2\Delta \frac{t_{l,\theta}+\Delta}{t_{l,\theta} + 2\Delta + t_l - 2(\gamma_m x_{\theta} + \gamma_n y_{\theta} + \gamma_{m,n}^z z_{\theta})}
\vspace{0cm}
\end{equation}
and we define
\begin{equation}\label{t l theta}
t_{j,\theta} = \sqrt{t_l^2 + 4|\gamma_{m,n}|^2 - 4t_l(\gamma_m x_{\theta} + \gamma_n y_{\theta} + \gamma_{m,n}^z z_{\theta}) }.
\end{equation}

Using our first assumption that $2(|\gamma_m|+|\gamma_n|+|\gamma_{m,n}^z|) \le t_l$ and the fact that $|x_{\theta}|,|y_{\theta}|,|z_{\theta}| \le 1$, we have:
\begin{align}
\vspace{-0.1cm}
t_l - 2(\gamma_m x_{\theta} + \gamma_n y_{\theta} + \gamma_{m,n}^z z_{\theta}) \nonumber \\*
 & \hspace{-1.5cm} \ge t_l - 2(|\gamma_m x_{\theta}|+|\gamma_n y_{\theta}| + |\gamma_{m,n}^z z_{\theta}|) \nonumber \\*
 & \hspace{-1.5cm} \ge t_l - 2(|\gamma_m|+|\gamma_n|+|\gamma_{m,n}^z|) \nonumber \\*
 & \hspace{-1.5cm} \ge 0. \nonumber
\vspace{-0.1cm}
\end{align}
Thus, we get $\Delta' \le 2\Delta$, and therefore $\tilde{h}(t;\theta_x,\theta_y) = 0$ for $t \notin [t_l,t_l+2\Delta)$. Let us write any $t \in [t_l,t_l+2\Delta)$ as $t=t_l+\eta$, with $0 \le \eta < 2\Delta$. Then
\begin{equation}\label{distorted pulse approximation 1}
\tilde{h}(t;\theta_x,\theta_y)=h(\tau_{m,n}(t_l+\eta;\theta_x,\theta_y) - \tau_{m,n}(t_l;\theta_x,\theta_y)).
\end{equation}
Using our second assumption that $\Delta \ll t_l$ and $\eta < 2\Delta$, we have $\eta \ll t_l$. We then approximate the argument of $h(\cdot)$ in \eqref{distorted pulse approximation 1} to first order:
\begin{equation}\label{distorted pulse approximation 2}
\tau_{m,n}(t_l+\eta;\theta_x,\theta_y) - \tau_{m,n}(t_l;\theta_x,\theta_y).
\end{equation}

To find the support explicitly, we expand the above inequality. For the left-hand side, we find that
\begin{equation}\label{distorted pulse support 2}
\tau_{m,n}(t;\theta_x,\theta_y) - \tau_{m,n}(t_l;\theta_x,\theta_y) = \sigma_{l,m,n}(\theta_x,\theta_y) + o(\eta^2),
\end{equation}
where
\begin{align}\label{distorted pulse support 3}
\vspace{-0.1cm}
\sigma_{l,m,n}(\theta_x,\theta_y) &= \nonumber \\*
  & \hspace{-1.5cm} \frac{1}{2} \left( 1 + \frac{t_l - 2(\gamma_m x_{\theta} + \gamma_n y_{\theta} + \gamma_{m,n}^z z_{\theta})}{\sqrt{t_l^2 - 4(\gamma_m x_{\theta} + \gamma_n y_{\theta} + \gamma_{m,n}^z z_{\theta})t_l + 4|\gamma_{m,n}|^2}} \right).
\vspace{-0.1cm}
\end{align}

We now extend our assumption that $2(|\gamma_m|+|\gamma_n|+|\gamma_{m,n}^z|) \le t_l$, and assume that $|\gamma_m|+|\gamma_n|+|\gamma_{m,n}^z| \ll t_l$. Hence, $|\gamma_{m,n}| = \sqrt{|\gamma_m|^2+|\gamma_n|^2+|\gamma_{m,n}^z|^2} < |\gamma_m|+|\gamma_n|+|\gamma_{m,n}^z| \ll t_l$. Using this assumption, we get $\sigma_{l,m,n}(\theta_x,\theta_y) \rightarrow 1$. Replacing $\eta = t - t_l$ and substituting back to \eqref{distorted pulse approximation 1}, results in
\begin{equation*}\label{distorted pulse approximation 4}
\vspace{-0.1cm}
\tilde{h}(t;\theta_x,\theta_y) \approx h(t - t_l), t \in [t_l, t_l + 2\Delta).
\end{equation*}
Substituting back to \eqref{distorted delays 2} and using the fact that $h(t-t_l) = 0$ for $t \notin [t_l, t_l + 2\Delta)$, we get:
\begin{equation}\label{distorted delays 3}
\hat{\varphi}_{m,n}(t;\theta_x,\theta_y) \approx \sum_{l=1}^L {\tilde{a}_{l,m,n} h(t-t_l)}.
\end{equation}
Finally, plugging this back into \eqref{phi beamformed},
\begin{align}\label{phi beamformed FRI proof}
\vspace{-0.1cm}
\Phi(t;\theta_x,\theta_y)&=\frac{1}{N_{\textrm{RX}}}\sum_{(m,n) \in \mathcal{M}}{\hat{\varphi}_{m,n}(t;\theta_x,\theta_y)} \nonumber \\*
&\approx \frac{1}{N_{\textrm{RX}}}\sum_{(m,n) \in \mathcal{M}}{\sum_{l=1}^L {\tilde{a}_{l,m,n} h(t-t_l)}} \nonumber \\*
&= \sum_{l=1}^L{\frac{1}{N_{\textrm{RX}}} \sum_{(m,n) \in \mathcal{M}}{\tilde{a}_{l,m,n}}h(t-t_l)} \nonumber \\*
&= \sum_{l=1}^L{\tilde{b}_l h(t-t_l)}.
\vspace{-0.1cm}
\end{align}

Thus, we have shown that the beamformed signal obeys the FRI model.

\section{Beamformed Signal Support}
\label{sec:app beamformed support}

We consider the FRI model for the individual signals in \eqref{individual FRI}. According to our second assumption in Appendix \ref{sec:app beamformed FRI}, $h(t)$ is the known pulse-shape with a support of $[0,\Delta)$ for some known $\Delta$ satisfying $\Delta \ll T$.

We neglect all reflections that reach the $(m,n)$ transducer at times greater than $T$, considering them as noise. Therefore, for all $1 \le l \le L$ and $(m,n) \in \mathcal{M}$:
\begin{equation}\label{delays bound 1}
\vspace{-0.1cm}
t_{l,m,n} + \Delta \le T.
\vspace{-0.1cm}
\end{equation}
Using \eqref{distorted delays}, \eqref{delays bound 1} and the fact that $\tau_{m,n}(t;\theta_x,\theta_y)$ is non-decreasing for $t \ge 0$, we get:
\begin{equation}\label{delays bound 2}
\vspace{-0.1cm}
t_l \le \tau_{m,n}^{-1}(T-\Delta;\theta_x,\theta_y),
\vspace{-0.1cm}
\end{equation}
with $\tau_{m,n}^{-1}(t;\theta_x,\theta_y)$ being the inverse of $\tau_{m,n}(t;\theta_x,\theta_y)$ with respect to $t$:
\begin{equation}\label{tau mn inverse}
\vspace{-0.1cm}
\tau_{m,n}^{-1}(t;\theta_x,\theta_y) = \frac{t^2 - |\gamma_{m,n}|^2}{t - (\gamma_m x_{\theta} + \gamma_n y_{\theta} + \gamma_{m,n}^z z_{\theta})},
\vspace{-0.1cm}
\end{equation}
for $t \ge |\gamma_{m,n}|$. Assuming the pulse-shape to have a negligible support with respect to the penetration depth, $\Delta \ll T$, and using the fact that \eqref{delays bound 2} holds for all $(m,n) \in \mathcal{M}$, we get, for all $1 \le l \le L$:
\begin{equation}\label{delays bound 3}
\vspace{-0.1cm}
t_l \le \min_{(m,n) \in \mathcal{M}} \tau_{m,n}^{-1}(T;\theta_x,\theta_y).
\vspace{-0.1cm}
\end{equation}
Since $\{t_l\}_{l=1}^L$ denote the arrival times of the echoes to the reference element, we can set the upper bound $T_B(\theta_x,\theta_y)$ on the beamformed signal as:
\begin{equation}\label{TB app}
\vspace{-0.1cm}
T_B(\theta_x,\theta_y) = \min_{(m,n) \in \mathcal{M}} \tau_{m,n}^{-1}(T;\theta_x,\theta_y).
\vspace{-0.1cm}
\end{equation}

We are now left to show that $T_B(\theta_x,\theta_y) < T$. This holds since we can always find an element $(m_1,n_1) \in \mathcal{M}$ such that $\gamma_{m_1}$ and $\gamma_{n_1}$ have opposite signs to that of $x_{\theta}$ and $y_{\theta}$, respectively. Furthermore, we note that we can always place the reference element $(m_0,n_0)$ such that $\gamma_{m_1,n_1}^z = 0$ for a specific choice of $(m_1,n_1) \in \mathcal{M}$. Thus:
\begin{align}\label{TB bound 1}
\vspace{-0.1cm}
T_B(\theta_x,\theta_y) &\le \tau_{m_1,n_1}^{-1}(T;\theta_x,\theta_y) \nonumber \\*
&= \frac{T^2 - |\gamma_{m_1,n_1}|^2}{T + |\gamma_{m_1} x_{\theta}| + |\gamma_{n_1} y_{\theta}|} \nonumber \\*
&\le \frac{T^2 - |\gamma_{m_1,n_1}|^2}{T} \nonumber \\*
&\le T.
\vspace{-0.1cm}
\end{align}
By applying $\tau_{m,n}(t;\theta_x,\theta_y)$ on both sides of \eqref{TB app}, we also have:
\begin{equation}\label{TB bound 2}
\vspace{-0.1cm}
\tau_{m,n}(T_B(\theta_x,\theta_y);\theta_x,\theta_y) \le T.
\vspace{-0.1cm}
\end{equation}
\end{appendices}
%


\vfill\pagebreak

\bibliographystyle{IEEEbib}
\bibliography{general}

\end{document}